\pdfoutput=1
\documentclass[11pt,reqno,preprint]{article}
\usepackage{jheppub,epsfig,amssymb,amsmath,mathrsfs,hyperref}
\usepackage{multirow,feynmp-auto,bbm}

\def\be{\begin{equation}}
\def\ee{\end{equation}}
\def\ba{\begin{eqnarray}}
\def\ea{\end{eqnarray}}
\newcommand{\bea}{\begin{eqnarray}}
\newcommand{\eea}{\end{eqnarray}}
\newcommand{\nn}{\nonumber}
\def\nl{\nonumber\\}
\def\Li{\textrm{Li}}

\def\e{\epsilon}

\newcommand{\de}{\delta}
\newcommand{\lnden}{\ln|\de|}

\newcommand{\lndene}[1]{\ln^{#1}|\de|}

\def\fig#1{fig.~{\ref{#1}}}

\def\eqn#1{eq.~(\ref{#1})}

\def\eqns#1#2{eqs.~(\ref{#1}) and~(\ref{#2})}

\def\eqn#1{eq.~(\ref{#1})}
\def\eqns#1#2{eqs.~(\ref{#1}) and~(\ref{#2})}

\def\Hhex{{\cal H}^{\rm hex}}

\newcommand{\fwboxL}[2]{\text{\makebox[#1][l]{$#2$}}}

\def\EE{\mathcal{E}}
\def\Et{\tilde{E}}

\def\Gcusp{\Gamma_{\rm cusp}}
\def\Gvirt{\Gamma_{\rm virt}}
\def\ws{{w^\ast}}

\newcommand{\cE}{\begin{cal}E\end{cal}}

\newcommand{\cN}{\begin{cal}N\end{cal}}
\newcommand{\cO}{\begin{cal}O\end{cal}}
\newcommand{\cP}{\begin{cal}P\end{cal}}

\newcommand{\cW}{\begin{cal}W\end{cal}}

\title{Six-Gluon Amplitudes
in Planar ${\cal N}=4$ Super-Yang-Mills Theory at Six and Seven Loops}

\author{Simon~Caron-Huot,$^1$}
\author{Lance~J.~Dixon,$^{2,3,4}$}
\author{Falko Dulat,$^{2}$}
\author{Matt~von~Hippel,$^{5,6}$}
\author{Andrew~J.~McLeod$^{2,3,6}$}
\author{and Georgios~Papathanasiou$^{3,7}$}

\affiliation{$^1$ Department of Physics, McGill University, 
3600 Rue University, Montr\'eal, QC Canada H3A 2T8}

\affiliation{$^2$ SLAC National Accelerator Laboratory,
Stanford University, Stanford, CA 94309, USA}

\affiliation{$^3$ Kavli Institute for Theoretical Physics, 
UC Santa Barbara, Santa Barbara, CA 93106, USA}

\affiliation{$^4$ Institut f\"ur Physik and IRIS Adlershof,
Humboldt-Universit\"at zu Berlin, \\
Zum Gro\ss en Windkanal 6, D-12489 Berlin, Germany}

\affiliation{$^5$ Perimeter Institute for Theoretical Physics, 
Waterloo, Ontario N2L 2Y5, Canada}

\affiliation{$^6$ Niels Bohr International Academy, Blegdamsvej 17,
2100 Copenhagen, Denmark}

\affiliation{$^7$ DESY Theory Group, DESY Hamburg, Notkestra{\ss}e 85,
D-22607 Hamburg, Germany}

\abstract{We compute the six-particle maximally-helicity-violating (MHV) and next-to-MHV (NMHV) amplitudes in planar maximally supersymmetric Yang-Mills theory through seven loops and six loops, respectively, as an application of the extended Steinmann relations and using the cosmic Galois coaction principle. Starting from a minimal space of functions constructed using these principles, we identify the amplitude by matching its symmetries and predicted behavior in various kinematic limits.  Through five loops, the MHV and NMHV amplitudes are uniquely determined using only the multi-Regge and leading collinear limits.  Beyond five loops, the MHV amplitude requires additional data from the kinematic expansion around the collinear limit, which we obtain from the Pentagon Operator Product Expansion, and in particular from its single-gluon bound state contribution.  We study the MHV amplitude in the self-crossing limit, where its singular terms agree with previous predictions. Analyzing and plotting the amplitudes along various kinematical lines, we continue to find remarkable stability between loop orders.}

\emailAdd{schuot@physics.mcgill.ca}\emailAdd{lance@slac.stanford.edu}\emailAdd{dulatf@slac.stanford.edu}\emailAdd{mvonhippel@nbi.ku.dk}\emailAdd{amcleod@nbi.ku.dk}\emailAdd{georgios.papathanasiou@desy.de}

\preprint{ \begin{flushright} DESY 19-042 \\ HU-EP-19/04 \\ SLAC--PUB--17413 \end{flushright}}

\begin{document}
\hypersetup{pageanchor=false}
\maketitle
\hypersetup{pageanchor=true}
\begin{fmffile}{feyndiags}

\section{Introduction}

Recent years have seen enormous progress in our understanding and ability to compute scattering amplitudes, particularly in the fertile testing ground of planar ${\cal N}=4$ super-Yang-Mills (SYM) theory~\cite{Brink:1976bc,Gliozzi:1976qd}. This progress has included advances in the construction of integrands that contribute to this theory~\cite{Bern:2006ew,Bern:2007ct,Bern:2008ap,ArkaniHamed:2009dn,ArkaniHamed:2009sx,ArkaniHamed:2010kv,Bourjaily:2011hi,ArkaniHamed:2012nw,Arkani-Hamed:2013jha,Arkani-Hamed:2013kca,Lipstein:2012vs,Lipstein:2013xra} as well as the amplitudes resulting from their integration~\cite{DelDuca:2009au,DelDuca:2010zg,Goncharov:2010jf,Gaiotto:2011dt,Dixon:2011pw,Dixon:2011nj,CaronHuot:2011kk,Golden:2013xva,Golden:2013lha,Dixon:2013eka,Golden:2014xqa,Dixon:2014voa,Dixon:2014xca,Golden:2014pua,Dixon:2014iba,Drummond:2014ffa,Dixon:2015iva,Dixon:2016apl,Caron-Huot:2016owq,Dixon:2016nkn}. In particular, infrared divergences in this theory are understood to all orders via the BDS ansatz~\cite{Bern:2005iz}. This ansatz completely describes amplitudes involving four or five particles, a fact which was later understood as a consequence of the theory's (anomalous) dual conformal invariance~\cite{Drummond:2006rz, Bern:2006ew, Bern:2007ct, Alday:2007hr, Bern:2008ap, Drummond:2008vq}. For six or more particles the ansatz is corrected \cite{Alday:2007he,Bartels:2008ce} by finite dual conformally invariant (DCI) contributions~\cite{Drummond:2006rz,Drummond:2007aua,Nguyen:2007ya,Bern:2008ap,Drummond:2008aq,Paulos:2012nu,Caron-Huot:2018dsv,Bourjaily:2018aeq}.

These DCI contributions to the amplitude have especially nice properties. In particular, they have been observed to have uniform transcendental weight equal to twice the loop order. While generic amplitudes in this theory may depend on elliptic polylogarithms~\cite{Paulos:2012nu,CaronHuot:2012ab,Bourjaily:2017bsb} and even more complicated~\cite{Bourjaily:2018ycu,Bourjaily:2018yfy} functions, maximally-helicity-violating (MHV) and next-to-MHV (NMHV) amplitudes are expected~\cite{ArkaniHamed:2012nw} to contain only functions drawn from the class of generalized polylogarithms. This class of functions is well understood, and has a Hopf algebra coaction structure that has been exploited to great effect~\cite{Gonch2,Gonch3,FBThesis,Goncharov:2010jf,Brown:2011ik,Duhr:2011zq,Duhr:2012fh}. (We will often refer to this coaction loosely as a coproduct, even though technically it is not because the two objects a function is mapped to under it are really of different types.) Maximally iterating this coproduct yields an object called the \textit{symbol} of a function~\cite{Goncharov:2010jf}.  In six- and seven-particle kinematics (and for higher multiplicity two-loop MHV amplitudes) the letters entering the symbol have intriguing connections to cluster algebras~\cite{Golden:2013xva,Golden:2013lha,Golden:2014pua,Golden:2014xqa,Harrington:2015bdt,Drummond:2017ssj,Drummond:2018dfd,Golden:2018gtk,Golden:2019kks}. 

Using these known symbol letters in six-particle kinematics and considering functions with physical branch cuts, it is possible to write down the space of functions within which the amplitude should reside. Moreover, by requiring a general ansatz of these functions to have the right symmetries and kinematic properties, the amplitude can be identified uniquely within this space. This bootstrap procedure was first employed for the three-loop six-particle MHV amplitude~\cite{Dixon:2011pw,Dixon:2013eka}.  Since then, it has been used to fix six-particle amplitudes through five loops~\cite{Dixon:2011nj, Dixon:2014voa, Dixon:2014xca, Dixon:2014iba, Dixon:2015iva, Caron-Huot:2016owq} and seven-particle amplitudes through four loops~\cite{Drummond:2014ffa, Dixon:2016nkn,Drummond:2018caf}.

Crucial to this progress has been an increasingly refined understanding of the space of functions needed to represent these amplitudes. In particular, the hexagon and heptagon bootstraps have in recent years been augmented by knowledge of the Steinmann relations~\cite{Steinmann,Steinmann2,Cahill:1973qp,Caron-Huot:2016owq,Dixon:2016nkn}. In a companion paper~\cite{CGG}, we present the next (and probably final) evolution of this space.  We find that the Steinmann relations can be applied at any depth in the symbol~\cite{Caron-Huot:2018dsv,CGG}.  These conditions apply at the level of the symbol, but there are also important restrictions on the function space that involve multiple zeta value (MZV) constants, which are invisible at the level of the symbol.  Once the amplitudes are properly normalized, they belong to a space of functions that satisfies a cosmic Galois coaction principle~\cite{Schnetz:2013hqa,Brown:2015fyf,Panzer:2016snt,Schnetz:2017bko}, which provides important restrictions on the MZVs that can appear.  Furthermore, the space of functions is {\it minimal} through weight seven:  it is impossible to eliminate any more functions, because all of them are needed to capture the iterated derivatives (or coproduct entries) of the six and seven loop amplitudes. We call this minimal space of functions $\Hhex$.

In this paper, we employ this space of functions to construct the six-particle amplitude through seven loops in the MHV sector and through six loops in the NMHV sector. After constructing an ansatz with the appropriate symmetries, we require that it behaves appropriately under the action of the dual superconformal $\bar{Q}$ operator~\cite{CaronHuot:2011kk}---which constrains the symbol letters that can appear in the final entry---and we require the expected leading-power behavior in the collinear limit. To do this, we construct a basis of functions iteratively in the weight, and work out their limiting behavior. At six and seven loops, these constraints leave fewer than 20 free parameters, out of thousands of initial parameters.

To fix the remaining parameters, we consider additional kinematic limits that supply independent information about the amplitude. We first consider the multi-Regge limit, in which outgoing particles are widely separated in rapidity, and the amplitude factorizes in a Fourier-Mellin transformed space~\cite{Bartels:2008ce,Bartels:2008sc,Fadin:2011we,Lipatov:2012gk,Dixon:2014iba}.  The behavior in this limit is now understood to all orders via integrability techniques~\cite{Basso:2014pla}.  Constraints from this limit yield a unique answer for the amplitude through five loops and for the six-loop NMHV amplitude. However, at both six and seven loops for the MHV amplitude we observe a novel qualitative feature: the appearance of a single function that vanishes in the leading-power collinear and multi-Regge limits.  Hence, the free parameter multiplying this function cannot be determined from these limits alone.

To determine this one remaining coefficient in our ansatz, we use a kinematic expansion around the collinear limit, which can be computed in the framework of the Wilson loop (or Pentagon) Operator Product Expansion (OPE)~\cite{Alday:2010ku,Basso:2013vsa,Basso:2013aha,Basso:2014koa,Basso:2014jfa,Basso:2014nra,Belitsky:2014sla,Belitsky:2014lta,Basso:2014hfa,Belitsky:2015efa,Basso:2015rta,Basso:2015uxa,Belitsky:2016vyq}. Quite interestingly, this last step requires going up to {\it second} order in the near-collinear expansion, and specifically examining the OPE contribution of the first gluon bound state.

Having determined the amplitudes, we proceed to study their properties, which reveals another novel feature that first appears at six loops: a previously conjectured cross-loop-order relation between MHV amplitudes and NMHV amplitudes~\cite{Dixon:2014iba,Caron-Huot:2016owq}, which held for several loop orders, no longer seems to hold.  As part of our analytic study, we present formulas for the values of our amplitudes at various points in the space of kinematics, in order to exhibit their number-theoretic properties. This study includes the limit of physical $2\to4$ and $3\to3$ kinematics where the corresponding Wilson loop approaches a self-crossing configuration~\cite{Georgiou:2009mp,Dorn:2011gf,Dorn:2011ec,Dixon:2016epj}, and the singular terms can be resummed to arbitrary loop order. We also plot the perturbative amplitudes numerically along various lines in the Euclidean region.  The new analytic features we described do not affect a remarkable numerical consistency between loop orders at generic values of the cross ratios.  We interpret this consistency as evidence that the perturbative expansion of the amplitudes in these regions has a finite radius of convergence.

This paper is organized as follows. In section \ref{sec:space_review}, we begin by reviewing our minimal space of functions, and discuss how to normalize the six-particle MHV and NMHV amplitudes in order to fit them into this space. Then, in section \ref{sec:constraints}, we describe how to construct the amplitude through seven loops in the MHV sector and at six loops in the NMHV sector, by applying constraints from symmetries and kinematic limits. Section \ref{sec:Numerics} contains the number-theoretic and numeric exploration of the amplitudes on a variety of kinematic points and lines.  In section \ref{sec:SelfCrossing} we examine the self-crossing limit of the MHV amplitude.  In section \ref{sec:Conclusions} we conclude and discuss directions for future research. 

We provide the following ancillary files along with this paper: 
\texttt{SixGluonAmpsAndCops.m},\\\noindent
\texttt{SixGluonHPLLines.m}, \texttt{SelfCross.m}, 
\texttt{SelfCrossSingular.m}, \texttt{hexMRKL1-7.m}, \texttt{WL0-6.m}, \newline\texttt{W1111L0-6.m}, \texttt{WL7.m},  and \texttt{WLOPEblocks.m}.
These computer-readable files describe results that are too lengthy to place in the text, including (respectively) a coproduct description of the amplitudes, their values on certain Euclidean lines in terms of harmonic polylogarithms~\cite{Remiddi:1999ew}, the MHV values in the self-crossing limit, a formula for the singular self-crossing behavior to 20 loops, formulas for the amplitudes in the multi-Regge limit, formulas for the near-collinear limit of the framed Wilson loop through six loops for MHV and NMHV and at seven loops for MHV, and individual contributions to the Wilson loop OPE. The files are hosted at~\cite{CosmicWebsite}.


\section{Review and normalization}
\label{sec:space_review}

\subsection{Superamplitudes, kinematic variables and generalized polylogarithms}

Let us begin by briefly recalling certain facts about the general structure of six-particle amplitudes in planar $\cN=4$ SYM theory, including their kinematic dependence and the class of functions that encompasses them. Due to the supersymmetry of this theory, we may combine color-ordered amplitudes with different external particles into corresponding superamplitudes, which are defined in an on-shell superspace~\cite{Nair:1988bq,Georgiou:2004by,Bianchi:2008pu,ArkaniHamed:2008gz}. The superfield $\Phi$ can be written in terms of Grassmann variables $\eta^A$ that transform in the fundamental representation of the $SU(4)$ R-symmetry,
\be
\Phi\ =\ G^+ + \eta^A \Gamma_A + \tfrac{1}{2!} \eta^A \eta^B S_{AB}
+ \tfrac{1}{3!} \eta^A \eta^B \eta^C \epsilon_{ABCD} \overline{\Gamma}^D
+ \tfrac{1}{4!} \eta^A \eta^B \eta^C \eta^D \epsilon_{ABCD} G^-,
\label{onshellmultiplet}
\ee
where the gluons of each helicity $G^\pm$, gluinos of each helicity $\Gamma_A$ and $\overline{\Gamma}^A$, and scalars $S_{AB}$ represent the on-shell particle content of the theory.

The superamplitude $\mathcal{A}_n(\Phi_1,\Phi_2,\ldots,\Phi_n)$ is typically broken into three factors: the BDS ansatz $\mathcal{A}_n^{\rm BDS}$~\cite{Bern:2005iz}, the remainder function ${\cal R}_n$~\cite{Bern:2008ap}, and a ratio function $\cP_n$~\cite{Drummond:2008vq}, giving
\be
\mathcal{A}_n\ =\ \mathcal{A}_n^{\rm BDS}\times\exp({\cal R}_n)\times\cP_n\,.
\label{Rndef}
\ee
The BDS ansatz encodes the infrared-divergent part of the amplitude (in dimensional regularization) as well as its non-DCI part. This leaves the remainder function and ratio function, which are finite and respect dual conformal symmetry. The ratio function carries all dependence on the Grassmann variables and can be expanded in Grassmann degree to isolate contributions with different helicity structure:
\be \label{eq:ratio_func_expansion}
\cP=1+\cP_{\rm NMHV}+\cP_{\rm N^2MHV}+\ldots+\cP_{\rm \overline{MHV}}\,.
\ee
The kinematic dependence of the remainder and ratio function is most conveniently written in terms of dual variables $x_i$ and $\theta_i$, which are defined in terms of the external momenta $k_i$ and the Grassmann variables $\eta^A$ via 
\be
k_i^{\alpha \dot\alpha} = 
\lambda_i^\alpha \tilde{\lambda}_i^{\dot\alpha}
= x_i^{\alpha \dot\alpha} - x_{i+1}^{\alpha \dot\alpha}, \qquad
\lambda_i^{\alpha} \eta_i^A
= \theta_i^{\alpha A} - \theta_{i+1}^{\alpha A}\,.
\label{xthetadef}
\ee
Here the additional index $i$ on the Grassmann variables associates them with the $i^\text{th}$ external particle, and $\lambda$, $\tilde \lambda$ are two-component spinors. For more background on these dual \mbox{(super-)}coordinates, see for instance ref.~\cite{Elvang:2013cua}.

Focusing now on the case of six particles, there are only three helicity configurations: MHV, NMHV, and $\overline{\rm MHV}$. The $\overline{\rm MHV}$ amplitude is parity conjugate to MHV, and thus the entire amplitude is encoded in the MHV and NMHV components. After normalizing them by the BDS ansatz, these amplitudes only depend on the kinematics through dual-conformal invariant cross ratios. Only three algebraically independent cross ratios can be formed for six particles, and traditionally they have been chosen as 
\be\label{uvw_def}
u = \frac{x_{13}^2\,x_{46}^2}{x_{14}^2\,x_{36}^2}\,, 
\qquad v = \frac{x_{24}^2\,x_{51}^2}{x_{25}^2\,x_{41}^2}\,, \qquad
w = \frac{x_{35}^2\,x_{62}^2}{x_{36}^2\,x_{52}^2}\, ,
\ee
where $x_{ij}^2 \equiv (x_i^\mu - x_j^\mu)^2$ are squared differences of dual coordinates. These cross ratios can also be expressed in terms of (planar) two- and three-particle Mandelstam invariants using the translation
$s_{i,i+1,\dots,i+n-1} = (k_i+k_{i+1}+\cdots+k_{i+n-1})^2 = x_{i,i+n}^2$.

The MHV amplitude corresponds to the leading term in the expansion~\eqref{eq:ratio_func_expansion}, and as such it depends only on the remainder function ${\cal R}_6(u,v,w)$ defined in \eqn{Rndef}. This function is expected to be a pure (generalized) polylogarithmic function (to be defined more precisely below) of the cross ratios~\eqref{uvw_def} to all loop orders, meaning that the kinematic dependence only appears in polylogarithms and not in any rational prefactors multiplying these functions. The NMHV contribution to the ratio function $\cP_{\rm NMHV}$ is not pure, but it can be written as a sum of pure polylogarithmic functions multiplied by $R$-invariants.

To define the $R$-invariants, we first recall the definition of momentum supertwistors~\cite{Hodges:2009hk,Mason:2009qx},
\be
\mathcal{Z}_i = (Z_i \, | \, \chi_i), \qquad 
Z_i^{R=\alpha,\dot\alpha} = 
(\lambda_i^\alpha , x_i^{\beta \dot\alpha}\lambda_{i\beta}),
\qquad
\chi_i^A= \theta_i^{\alpha A}\lambda_{i \alpha} \,.
\ee
The momentum twistors $Z_a$, considered as vectors in $\mathbb{CP}^3$, can be contracted into SL(4) invariants using the Levi-Civita tensor,
\be\label{eq:fourbrak}
\langle abcd\rangle \equiv \epsilon_{RSTU} Z_a^R Z_b^S Z_c^T Z_d^U\,,
\ee
where in particular $x^2_{ij}\propto \langle i-1ij-1j\rangle$, and the additional factors (which would make this proportionality exact) cancel out in the fully dual-conformal invariant ratios in~\eqn{uvw_def}. Similarly, five-brackets of the $\mathcal{Z}_a$ are dual superconformal invariants known as $R$-invariants. They are defined to be 
\be
(f)\ \equiv\ [abcde]\ =\ 
\frac{\delta^4\bigl(\chi_a \langle bcde\rangle + {\rm cyclic}\bigr)}
{\langle abcd\rangle\langle bcde\rangle
\langle cdea\rangle\langle deab\rangle\langle eabc\rangle}\,,
\label{five_bracket_def}
\ee
where we denote the five-bracket of legs $\{a,b,c,d,e\}$ by the remaining leg $f$. Using these quantities, we can parametrize $\cP_{\rm NMHV}$ in terms of a parity-even function $V$ and a parity-odd function $\tilde{V}$:
\begin{align}
&\cP_{\rm NMHV}\ =\ \frac{1}{2}\Bigl[
 [(1) + (4)] V(u,v,w) + [(2) + (5)] V(v,w,u) + [(3) + (6)] V(w,u,v) 
\nonumber \\
&\hskip2.2cm 
+ [(1) - (4)] \tilde{V}(u,v,w) - [(2)-(5)] \tilde{V}(v,w,u)
  + [(3) - (6)] \tilde{V}(w,u,v) \Bigr] \,. \ \ \ 
\label{PVform}
\end{align}

The remainder function, $V$, and $\tilde{V}$ have perturbative expansions in the large-$N$ coupling $g^2$ defined by
\be
g^2 = \frac{\lambda}{16\pi^2} = \frac{N g_{\rm YM}^2}{16\pi^2}\,,
\ee
where $\lambda$ is the usual 't Hooft coupling.
The coefficients of the perturbative expansion are linear combinations of generalized polylogarithms. These functions are defined as iterated integrals over logarithmic kernels, commonly denoted by 
\be \label{eq:G_func_def}
G_{a_1,\dots, a_n}(z) = \int_0^z \frac{dt}{t-a_1} G_{a_2,\dots, a_n}(t)\,, \qquad G_{\fwboxL{27pt}{{\underbrace{0,\dots,0}_{p}}}}(z) = \frac{\ln^p z}{p!} \, , 
\ee
with the recursion starting at $G(z)=1$.
The number $n$ of nested integrations in the above definition is referred to as the \emph{weight} of the generalized polylogarithm.

At $L$ loops in the perturbative expansion in $g^2$, the remainder function, $V$, and $\tilde{V}$ and are all pure functions of uniform transcendental weight $2L$. In practice this means that the total differential of these functions, and the functions $F$ we will be interested in later, can be written in the form 
\be \label{eq:A_differential}
d F
= \sum_{s \in \cal{S}_\text{hex}} F^{s} \ d \ln s \, ,
\ee
where ${\cal S}_{\text{hex}}$ is the set of nine hexagon symbol letters that will be introduced below. If $F$ has weight $n$, then each $F^{s}$ is a pure function of weight $n-1$.

Because of the motivic structure of generalized polylogarithms~\cite{Gonch2,Goncharov:2010jf,Brown:2011ik,Duhr:2012fh,Brown1102.1312,Brown:2015fyf}, the total differential~\eqref{eq:A_differential} corresponds to the component of the coaction with weight-one functions in its back entry, 
\be
\Delta_{n-1,1} (F )
\ =\ \sum_{s \in \cal{S}_\text{hex}} F^{s} \otimes \ln s \,.
\ee
The map $\Delta_{\bullet,1}$ can be applied iteratively to the functions in the left factor of the coproduct, breaking down the weight $n$ polylogarithms entering $F$ into $n$-fold tensor products of logarithms, objects commonly referred to as symbols~\cite{Goncharov:2010jf}. The letters $s$ appearing in the arguments of the logarithms belong to the hexagon {\it symbol alphabet}:
\be
s \in {\cal S}_\text{hex} =
\left\{u, v, w, 1-u, 1-v, 1-w, y_u, y_v, y_w \right\} \,.
\label{hex_letters_app} 
\ee
Here we have introduced three parity-odd letters,
\be
y_u = \frac{u-z_+}{u-z_-}\,, \qquad y_v = \frac{v-z_+}{v-z_-}\,,
\qquad y_w = \frac{w - z_+}{w - z_-}\, ,
\label{yfromu_app}
\ee
defined in terms of the quantities
\be
z_\pm = \frac{1}{2}\Bigl[-1+u+v+w \pm \sqrt{\Delta}\Bigr],
\qquad
\Delta = (1-u-v-w)^2 - 4 u v w .
\ee
(The cross ratios~\eqref{uvw_def} are parity even.) 

In this paper we will often use the following equivalent symbol alphabet,
\be
{\cal S}^{\prime}_\text{hex} =
\left\{a, b, c, m_u, m_v, m_w, y_u, y_v, y_w \right\} \,,
\label{hex_letters_abc_app}
\ee
which is related to the original alphabet ${\cal S_{\text{hex}}}$ through the relations
\be
a = \frac{u}{vw} \,, \quad
b = \frac{v}{wu} \,, \quad
c = \frac{w}{uv} \,, \quad
m_u = \frac{1-u}{u} \,, \quad
m_v = \frac{1-v}{v} \,, \quad
m_w = \frac{1-w}{w} \,.
\label{abcdef}
\ee
This alphabet makes the Steinmann relations more transparent by isolating the three independent three-particle Mandelstam invariants $s_{i,i+1,i+2}$ in different letters, $a$, $b$ and $c$.  In this respect it resembles more closely the usual choice of alphabet for the seven-particle amplitude~\cite{Drummond:2014ffa}.

At higher loops, expressing these functions in the $G$-function notation~\eqref{eq:G_func_def} becomes overly cumbersome, and it proves more effective to encode the amplitude in terms of its $\Delta_{\bullet, 1}$ coproduct. This amounts to keeping track of the (iterated) derivatives of the amplitude, as well as the integration constants required to reconstruct the amplitude from these derivatives. This way of treating these functions will be reviewed in more depth in a companion paper~\cite{CGG}, and has been discussed elsewhere in the literature (see for example ref.~\cite{Dixon:2013eka}). In the next subsection, we describe how we normalize the amplitudes so that they lie in $\Hhex$.

\subsection{Cosmic normalization}
\label{cosmicnorm}

In ref.~\cite{CGG} we describe the construction of a minimal space
of (extended) Steinman-satisfying hexagon functions. 
In addition to the extended Steinmann relations, this space is constructed to obey a coaction principle, which is to say it is invariant under the cosmic Galois
group~\cite{Schnetz:2013hqa,Brown:2015fyf,Panzer:2016snt,Schnetz:2017bko}. 
This condition implies that if one studies the coaction of higher-loop amplitudes, the first entry belongs to a stable space whose dimension at fixed weight saturates as the loop order increases. 
Empirically, this saturation happens around weight $L$, namely at around half the weight of the corresponding $L$-loop amplitude. While some of these restrictions have long been understood at symbol level, at the level of functions they result in further restrictions on the transcendental constants that are allowed to appear as free elements:
only $\zeta_4$, $\zeta_6$, $\zeta_8$, etc.~are needed.

It is highly nontrivial that the six-point amplitudes lie within this space through seven loops.  In fact, they only do so once they are properly normalized. First of all, in order to preserve the Steinmann constraints, we must normalize the amplitude by the BDS-like ansatz~\cite{Alday:2009dv,Dixon:2015iva,Caron-Huot:2016owq} (up to a kinematically constant factor).
The BDS-like ansatz differs from the BDS ansatz by a factor of
\be
\exp\biggl[ - \frac14 \Gcusp(g^2) \, \EE^{(1)}(u,v,w) \biggr] \, ,
\label{BVDSvsBDSlike}
\ee
where the cusp anomalous dimension is~\cite{Beisert:2006ez}
\be
\frac14 \Gcusp(g^2)\ =\ g^2 - 2\,\zeta_2\,g^4 + 22\,\zeta_4\,g^6
- \Bigl[ 219\, \zeta_6 + 8 \, (\zeta_3)^2 \Bigr] \, g^8 + \cdots,
\label{Gcusp}
\ee
and 
\be
\EE^{(1)}(u,v,w) = \Li_2\Bigl(1-\frac{1}{u}\Bigr)
+ \Li_2\Bigl(1-\frac{1}{v}\Bigr) + \Li_2\Bigl(1-\frac{1}{w}\Bigr)
\label{E1}
\ee
is the finite, dual conformally invariant part of the one-loop MHV
amplitude.

However, beyond two loops this normalization is not sufficient to place
the amplitudes into $\Hhex$, and needs to be
adjusted by a kinematical constant.  (The Steinmann relations uniquely fix
the normalization, up to this constant.)
One way to see the need for a new normalization is to inspect
the three-loop values of the MHV amplitude $\EE^{\rm old}$ and NMHV amplitude
$E^{\rm old}$ defined in ref.~\cite{Caron-Huot:2016owq}
at the point $(u,v,w)=(1,1,1)$:
\be
\EE^{{\rm old}\, (3)}(1,1,1) = \frac{413}{3} \, \zeta_6 + 8 (\zeta_3)^2 \,,
\qquad
E^{{{\rm old}\, (3)}}(1,1,1) = - \frac{940}{3} \zeta_6 + 8 (\zeta_3)^2 \,.
\label{old3loops111}
\ee
Both amplitudes contain $(\zeta_3)^2$ with the same nonzero coefficient.
The presence of $(\zeta_3)^2$ would violate the coaction principle,
because of a specific term in its coaction, namely
$\Delta_{3,3}[ (\zeta_3)^2] = 2 \zeta_3 \otimes \zeta_3$.
This term contains a $\zeta_3$ in its first entry,
but no $\zeta_3$ is allowed there, because the weight-three
functions in $\Hhex$ all vanish at
$(1,1,1)$~\cite{CGG}.

Therefore, starting at three loops,
we modify the BDS-like ansatz ${\cal A}_6^{\rm BDS-like}$
by a function of the coupling $\rho(g^2)$, in order to fit the amplitudes
into a space that is invariant under the cosmic Galois group.
In other words, we define the ``cosmically normalized'' MHV amplitude $\EE$ in
terms of the full, infrared-divergent amplitude ${\cal A}_6$, the
BDS-like ansatz, and $\rho$, as
\be
{\cal A}_6^{\rm MHV}(s_{ij},\e) =
{\cal A}_6^{\rm BDS-like}(s_{i,i+1},\e)
\times \rho(g^2) \times \EE(u,v,w,g^2).
\label{EXMHVdef}
\ee
The BDS-like ansatz is~\cite{Alday:2009dv,Dixon:2015iva}
\be
{\cal A}_6^{\rm BDS-like}(s_{i,i+1},\e)
= A_6^{\rm MHV,tree} 
 \exp\Biggl[ \sum_{L=1}^\infty (g^2)^L
            \Bigl( f^{(L)}(\e) \hat{M}_6(L\e) + C^{(L)} \Bigr) \Biggr] \,,
\label{BDSlikedef}
\ee
where $C^{(L)}$ is a constant at each loop order, as is $f^{(L)}(\e)$,
\be
f(\e)\ =\ \sum_{L=1}^\infty (g^2)^L f^{(L)}(\e)
     \ =\ \frac{1}{4} \Gcusp + {\cal O}(\e).
\label{fLdef}
\ee
(We suppress the two additional terms in the $\e$ expansion of $f$ here
for simplicity.)

Finally, the quantity $\hat{M}_6(\e)$ differs from
the full one-loop MHV amplitude by an amount proportional to $\EE^{(1)}$.
It is given by
\begin{align}
\hat{M}_6(\e) =&( 4 \pi e^{-\gamma_E} )^\e \sum_{i=1}^6  \biggl[ 
- \frac{1}{\e^2} \left( 1 + \e \ln \left( \frac{\mu^2}{-s_{i,i+1}} \right) + \frac{\e^2}{2} \ln^2 \left( \frac{\mu^2}{-s_{i,i+1}} \right) \right) \nonumber \\
&\qquad \qquad \qquad \qquad + \frac{1}{2} \ln^2 \left( \frac{s_{i,i+1}}{s_{i+1,i+2}} \right)- \frac{1}{4} \ln^2 \left( \frac{s_{i,i+1}}{s_{i+3,i+4}} \right) + \frac{3}{2} \zeta_2 \biggr]  +\cO(\epsilon) \,,
\label{Mhatdef}
\end{align}
where $\gamma_E$ is the Euler-Mascheroni constant.

It is important that $\hat{M}_6(\e)$ only depends on the two-particle
invariants $s_{i,i+1}$ to $\cO(\epsilon^0)$, so that factoring it out of the amplitude
does not affect the Steinmann relations for three-particle
invariants~\cite{Caron-Huot:2016owq}.

The relation between $\EE$ and the remainder function ${\cal R}_6$ is
\be
\EE = \frac{1}{\rho} \, 
\exp\biggl[ \frac{1}{4} \Gcusp \EE^{(1)} + {\cal R}_6 \biggr] \,.
\label{EXMHVtoR6}
\ee
The coefficient functions that specify the cosmically normalized NMHV amplitude, which we call $E$ and $\tilde E$, are defined similarly;
in terms of the scheme-independent coefficient functions $V$
 and $\tilde{V}$ of the ratio function \eqref{PVform} they are defined
by
\be
E = \EE \times V, \qquad
\Et = \EE \times \tilde{V}.
\label{EXNMHVdef}
\ee
To pass from the old normalization for
$\EE^{\rm old}$, $E^{\rm old}$, $\Et^{\rm old}$ used in
ref.~\cite{Caron-Huot:2016owq} to the one used here, simply
divide the old functions by $\rho(g^2)$.

We compute $\rho$ by requiring that the amplitudes fit into the minimal
space that obeys the coaction principle.  For example, at three loops we need to have
$\rho(g^2) = 1 + 8 (\zeta_3)^2 \, g^6 + {\cal O}(g^8)$ in order
to cancel the two appearances of $(\zeta_3)^2$ in \eqn{old3loops111}.
This criterion does not necessarily determine $\rho$ uniquely.
For instance, $\rho$ could be adjusted by terms involving even zeta values
$\zeta_{2L}$ at $L$ loops since these have free parameters associated with them
(except for $\zeta_2$ at one loop).  We choose to fix this ambiguity by not
including any $\zeta_{2L}$ term at $L$ loops.  Through seven loops, we find
that the minimal solution for $\rho$ does not require any genuine MZVs,
nor does it contain any factor of $\zeta_2$:
\bea
\rho(g^2) &=& 1 + 8 (\zeta_3)^2 \, g^6  - 160 \zeta_3 \zeta_5 \, g^8
+ \Bigl[ 1680 \zeta_3 \zeta_7 + 912 (\zeta_5)^2 - 32 \zeta_4 (\zeta_3)^2 \Bigr]
\, g^{10}
\nonumber\\
&&\null\hskip0.0cm
- \Bigl[ 18816 \zeta_3 \zeta_9 + 20832 \zeta_5 \zeta_7
  - 448 \zeta_4 \zeta_3 \zeta_5 - 400 \zeta_6 (\zeta_3)^2 \Bigr] \, g^{12}
\nonumber\\
&&\null\hskip0.0cm
+ \Bigl[ 221760 \zeta_3 \zeta_{11} + 247296 \zeta_5 \zeta_9 + 126240 (\zeta_7)^2
   - 3360 \zeta_4 \zeta_3 \zeta_7 - 1824 \zeta_4 (\zeta_5)^2
	\nonumber\\
&&\null\hskip0.7cm
 - 5440 \zeta_6 \zeta_3 \zeta_5 - 4480 \zeta_8 (\zeta_3)^2 \Bigr] \, g^{14}
\ +\ {\cal O}(g^{16}).
\label{rho}
\eea
This form for $\rho$ is uniquely fixed, given the following assumptions:
\begin{enumerate}
\item $\rho$ does not contain MZVs of depth two or higher, e.g.~no $\zeta_{5,3}$,
\item the coaction principle is satisfied through weight 14 at the point
$(1,1,1)$,
\item it is also satisfied at an analytic continuation of this point to
$3\to3$ self-crossing kinematics,
\item a subspace of the hexagon functions that we can define to all
weights, which saturates the space of MZVs in $\Hhex$ through weight 10, also does so at weight 11.
\end{enumerate}
The last constraint is only needed to uniquely fix $\rho^{(7)}$.  It 
imposes one additional constraint on the MZVs appearing at weight 11,
which in turn implies one fewer allowed MZV at weight 14.
Otherwise we would have an ambiguity in $\rho$ at seven loops, because
we do not yet know the NMHV amplitude at this order.

Once we have fixed $\rho$ at a given loop order, many constraints ensue
as we go to higher loop orders.  For example, none of the $\{6,1,1\}$
coproducts of the four-loop amplitudes can contain a $(\zeta_3)^2$
when evaluated at $(1,1,1)$. It is interesting that the same value of $\rho$ works 
for both the MHV and NMHV amplitudes. Perhaps this fact indicates that $\rho$ 
is determining a particular infrared regularization scheme.

In fact, $\rho$ seems to be related to the inverse of the
cusp anomalous dimension at low loop orders.\footnote{We thank
Mark Spradlin for discussions on this point.}
More precisely, the quantity $\rho - 4 \, g^2/\Gcusp$ contains only even
zeta values until five loops:
\bea
\rho - \frac{4\,g^2}{\Gcusp} &=& - 2 \, \zeta_2 \, g^2
+ 12 \, \zeta_4 \, g^4 - 100 \, \zeta_6 \, g^6
+ 994 \, \zeta_8 \, g^8
\nonumber\\
&&\null\hskip0.0cm
- \Bigl[ 10980 \, \zeta_{10} - 96 \, (\zeta_5)^2
       - 16 \, \zeta_4 \, (\zeta_3)^2 \Bigr] \, g^{10}
\nonumber\\
&&\null\hskip0.0cm
+ \Bigl[ \frac{89756216}{691} \, \zeta_{12} - 3360 \, \zeta_5 \, \zeta_7
       - 320 \, \zeta_4 \, \zeta_3 \, \zeta_5
       - 240 \, \zeta_6 \, (\zeta_3)^2 \Bigr] \, g^{12}
\nonumber\\
&&\null\hskip0.0cm
- \Bigl[ 1611350 \zeta_{14} - 30960 (\zeta_7)^2 - 48384 \zeta_5 \zeta_9
  - 1056 \zeta_4 (\zeta_5)^2
\nonumber\\
&&\null\hskip0.7cm
  - 3360 \zeta_4 \zeta_3 \zeta_7 - 4800 \zeta_6 \zeta_3 \zeta_5
  - 3024 \zeta_8 (\zeta_3)^2 \Bigr] \, g^{14}
\ +\ {\cal O}(g^{16}).
\label{rhominusinvGcusp}
\eea
Recall that the $g^4$, $g^6$ and $g^8$ terms could be removed from this
relation by redefining $\rho$, using the fact that the corresponding
zeta values are independent, constant elements of $\Hhex$.
Given how the cusp anomalous dimension is tied to infrared divergences, eq.~\eqref{rhominusinvGcusp} might provide a clue for how $\rho$ defines a ``cosmically-preferred'' infrared regularization scheme.

Through five loops,
the values of the MHV amplitude at $(1,1,1)$ are, in the new normalization,
\bea
\EE^{(1)}(1,1,1) &=& 0 \,,
\label{EXMHVg1_111}\\
\EE^{(2)}(1,1,1) &=& - 10 \, \zeta_4 \,,
\label{EXMHVg2_111}\\
\EE^{(3)}(1,1,1) &=& \frac{413}{3} \, \zeta_6 \,,
\label{EXMHVg3_111}\\
\EE^{(4)}(1,1,1) &=&  - \frac{5477}{3} \, \zeta_8
+ 24 \, \Bigl[ \zeta_{5,3} + 5 \, \zeta_3 \, \zeta_5 - \zeta_2 \, (\zeta_3)^2
\Bigr] \,,
\label{EXMHVg4_111}\\
\EE^{(5)}(1,1,1) &=& \frac{379957}{15} \, \zeta_{10}
- 12 \, \Bigl[ 4 \, \zeta_2 \, \zeta_{5,3} + 25 \, (\zeta_5)^2 \Bigr]
\nonumber\\
&&\null\hskip0.0cm
- 96 \, \Bigl[ 2 \, \zeta_{7,3} + 28 \, \zeta_3 \, \zeta_7
             + 11 \, (\zeta_5)^2 - 4 \, \zeta_2 \, \zeta_3 \, \zeta_5
             - 6 \, \zeta_4 \, (\zeta_3)^2 \Bigr] \,.
\label{EXMHVg5_111}
\eea
For the NMHV amplitude (parity even part) they are
\bea
E^{(1)}(1,1,1) &=& - 2 \, \zeta_2 \,,
\label{EXg1_111}\\
E^{(2)}(1,1,1) &=& 26 \, \zeta_4 \,,
\label{EXg2_111}\\
E^{(3)}(1,1,1) &=& - \frac{940}{3} \, \zeta_6 \,,
\label{EXg3_111}\\
E^{(4)}(1,1,1) &=&   \frac{36271}{9} \, \zeta_8
- 24 \, \Bigl[ \zeta_{5,3} + 5 \, \zeta_3 \, \zeta_5 - \zeta_2 \, (\zeta_3)^2
\Bigr] \,,
\label{EXg4_111}\\
E^{(5)}(1,1,1) &=& - \frac{1666501}{30} \, \zeta_{10}
+ 12 \, \Bigl[ 4 \, \zeta_2 \, \zeta_{5,3} + 25 \, (\zeta_5)^2 \Bigr]
\nonumber\\
&&\null\hskip0.0cm
+ 132 \, \Bigl[ 2 \, \zeta_{7,3} + 28 \, \zeta_3 \, \zeta_7
             + 11 \, (\zeta_5)^2 - 4 \, \zeta_2 \, \zeta_3 \, \zeta_5
             - 6 \, \zeta_4 \, (\zeta_3)^2 \Bigr] \,.
\label{EXg5_111}
\eea
We will give the six- and seven-loop values in section~\ref{SubsectionPoint111}.


\section{Bootstrapping the six-particle amplitude}
\label{sec:constraints}

In this section we describe the constraints that we impose in order to
uniquely determine the MHV amplitude through seven loops, and the NMHV
amplitude through six loops.

\subsection{Discrete symmetries, \texorpdfstring{$\bar Q$}{Qbar} supersymmetry, and collinear limit}
\label{subsec:initial}

Our starting point is the space of functions described in ref.~\cite{CGG}: generalized polylogarithms with symbol letters drawn from the alphabet ${\cal S}_\text{hex}$ of eq.~\eqref{hex_letters_app}, that have branch cuts ending where the cross ratios $u$, $v$, and $w$ become zero or infinity, and that satisfy the extended Steinmann relations and obey a cosmic Galois coaction principle.

The general linear combination of all such functions with transcendental weight $2L$ forms an initial ansatz for each of the (cosmically normalized) functions $\EE^{(L)}$, $E^{(L)}$, and $\Et^{(L)}$. These functions inherit a set of discrete symmetries from the dihedral symmetry of the full superamplitude: $\EE$ is fully symmetric under all permutations of $u,v,w$, while
\be
E(u,v,w)=E(w,v,u)\,,\quad \Et(y_u,y_v,y_w)=-\Et(y_w,y_v,y_u) \, .
\ee
We also impose the condition~\cite{Dixon:2014iba}
\be
\Et(y_u,y_v,y_w)+\Et(y_v,y_w,y_u)+\Et(y_w,y_u,y_v)=0\,,
\ee
which removes an unphysical degree of freedom in the function $\Et$, as any cyclically symmetric piece of $\Et$ drops out of the full amplitude due to the following linear relation between $R$-invariants,
\be
(1)+(3)+(5)\ =\ (2)+(4)+(6).
\ee

At this early stage it is convenient to impose the ``final entry condition'' that follows from the $\bar{Q}$ equation~\cite{CaronHuot:2011ky,Bullimore:2011kg,CaronHuot:2011kk}.  The dual superconformal generator $\bar{Q}$ is a first-order differential operator acting on the $n$-point $L$-loop N${}^k$MHV amplitude, which relates it to an integral over the $(n+1)$-point $(L-1)$-loop N$^{k+1}$MHV amplitude. By choosing appropriate differentials, the latter ``source term'' can be made to vanish, leading to a set of homogeneous constraints.  In the MHV case, they take the form,
\be
\EE^u + \EE^{1-u} = \EE^v + \EE^{1-v} = \EE^w + \EE^{1-w} = 0 \, .
\label{MHVFEasconstraint}
\ee
These relations imply that the final entries of $\EE$ can only be drawn from a
subset of the normal symbol letters, which we may write in the alphabet \eqref{hex_letters_abc_app} as
\be
\EE\, \text{final entries} \in \left\{m_u, m_v, m_w, y_u, y_v, y_w \right\}\,.
\label{MHVFE}
\ee
The NMHV final entry conditions are slightly more involved~\cite{Dixon:2015iva}. The allowed final symbol entries here depend on which $R$-invariant the polylogarithm multiplies. The allowed combinations are
\bea
&&(1) \, d\ln b \,,\quad (1) \, d \ln\biggl(\frac{m_w}{y_v m_u}\biggr) \,,
\label{dlnfinalentries}\\
&&\Bigl[ (2) + (5) + (3) + (6) \Bigr] d \ln m_v
+ (1) \, d\ln(m_w y_u)
+ (4) \, d \ln(m_u y_w) \,,
\nonumber
\eea
along with their cyclic images. Note that the appearance of products like $d\ln(m_w y_u)$ means that these constraints link $E$ and $\Et$; after imposing these final entry conditions, our ans\"{a}tze for these functions can no longer be considered independent.

We apply these constraints first because they are easy to impose, and because they constrain a large number of coefficients. For example, at six loops, of the 3692 undetermined coefficients present in our initial weight 12 ansatz, only 236 and 102 parameters remain in the MHV and NMHV amplitudes, respectively, after imposing the above symmetries and final entry conditions.

We next turn to the collinear limit, where the six-point amplitude reduces to a five-point amplitude times a universal splitting function. Since the six-point BDS ansatz captures the correct collinear limits of the six-point amplitude, both the remainder function and the ratio function must vanish at loop level in this limit. $\EE$ and $E$ do not vanish, but their behavior can be easily found by taking the collinear limit of~\eqn{EXMHVtoR6}. In the $u$, $v$, and $w$ variables, the collinear limit corresponds to $v\to0$ and $u\to1-w$, where we have that
\be
\EE \big|_{\rm coll}\ =\ \frac{1}{2}\left(E(v,w,u)+E(w,u,v)\right)\big|_{\rm coll}
\ =\ \frac{\exp\Bigl[\tfrac14 \Gcusp \, \EE^{(1)}|_{\rm coll}\Bigr]}{\rho} \,.
\ee
Similar constraints apply in the cyclically related collinear limits, but are automatically enforced by the discrete symmetry conditions already imposed.

Imposing this behavior in the strict collinear limit (which is to say, at leading power) fixes many of the remaining parameters in our ansatz.  As we will see in the next section, it leaves us with fewer than 20 free parameters to constrain.

\subsection{Multi-Regge kinematics}
\label{subsec:multi-regge}

Next we constrain these remaining parameters, using input from additional kinematic limits. The multi-Regge limit, in which the collision of two highly energetic particles produces four particles that are strongly ordered in rapidity, is a prime source of such input.  In terms of Mandelstam invariants, with particles 3 and 6 incoming and particles 1,2,4,5 outgoing, the $2 \to 4$ multi-Regge limit reads
\be
s_{12} \gg s_{345},s_{123} \gg s_{34},s_{45},s_{56} \gg s_{23},s_{61},s_{234} \,.
\label{mominvstrongorder}
\ee
In this section, we will use $(u_1,u_2,u_3)$ instead of $(u,v,w)$ to avoid confusion with the multi-Regge variable $w$. In the cross ratios the multi-Regge limit becomes,
\be \label{eq:multi_regge_cross_ratios}
u_1 \to 1\,,\qquad u_2,u_3 \to 0\,,
\ee
with the ratios
\be
\frac{u_2}{1-u_1}\equiv \frac{1}{(1+w)\,(1+\ws)} {\rm~~~~and~~~~} 
\frac{u_3}{1-u_1}\equiv \frac{w\ws}{(1+w)\,(1+\ws)}
\label{wdef}
\ee
held fixed in terms of the complex-valued multi-Regge variable $w$. The behavior of the remaining letters,
\be 
y_1 \to 1, \qquad y_2 \to \frac{1+w^*}{1+w} \,,  \qquad
y_3 \to \frac{(1+w)w^*}{w(1+w^*)} \,,
\label{yinMRK}
\ee
may be inferred from the following momentum twistor parametrization,
\be
\left(\begin{array}{cccccc}
 \frac{\sqrt{\tau}}{\sqrt{w}} & 1 & -1 & 0 & 0 & 0 \\
 0 & 0 & 0 & 1 & 1 & \frac{\sqrt{w^*}}{\sqrt{\tau}} \\
 \frac{1}{\sqrt{\tau} \sqrt{w^*}} & 0 & 0 & -1 & 0 & \frac{1}{\sqrt{\tau} \sqrt{w^*}} \\
 \sqrt{\tau} \sqrt{w} & 0 & 1 & 1 & 0 & 0 \\
\end{array}\right) \,,
\label{yMRK}
\ee
where the columns correspond to $(Z_1,\ldots,Z_6)$, the rows correspond to their components, and the limit amounts to $\sqrt{u_2 u_3}\simeq\tau\to0$.

In the Euclidean region, loop corrections to the BDS-normalized amplitude vanish in the multi-Regge limit~\cite{Bartels:2008ce}, due to its conformal equivalence to a soft limit~\cite{Dixon:2013eka,DelDuca:2016lad}. Nontrivial behavior in the limit is obtained by analytically continuing into physical $2\to4$ Minkowski kinematics.  For particles 3 and 6 in the initial state, the Mandelstam variables $s_{12}$ and $s_{45}$ in the numerator of $u_1$ should be continued into the time-like region, which amounts to continuing
\be\label{eq:MRK_continuation}
u_1\to e^{-2\pi i}u_1\,
\ee
before taking the limit.

As we will review shortly, it is possible to obtain direct predictions for the behavior of the amplitude in multi-Regge kinematics when the external particles are gluons. For the bosonic MHV amplitude this is true by definition, whereas for the NMHV superamplitude, we need to specialize to its gluonic components. Without loss of generality we may choose the $(\chi_4)^4$ component of the ratio function (that is, $\cP^{(4444)}_{\textrm{NMHV}}$) \cite{Dixon:2015iva}, which describes the NMHV helicity configuration\footnote{In conventions where the momenta $p_3,p_6$ are incoming, and the rest outgoing. In the limit, helicity is preserved along the lines of the incoming gluons, and the only other inequivalent helicity configuration may be obtained by a parity transformation.}
\be
3^+ 6^+ \, \to \, 2^+ 4^- 5^+ 1^+ \,.
\label{NMHVconfig}
\ee
Taking the limits of the $R$-invariants that the functions $E$ and $\tilde E$ multiply, we may express the corresponding BDS-normalized gluon amplitude in multi-Regge kinematics as
\begin{align}
\cP^{(4444)}_{\rm NMHV}\, e^{{\cal R}_6}\ =& \frac{\rho e^{-\frac{\Gamma_{\text{cusp}}}{4}\cE^{(1)}}}{2(1+w^*)}
\Big\{ E(u_1,u_2,u_3)+E(u_3,u_1,u_2)+\tilde{E}(y_1,y_2,y_3)-\tilde{E}(y_3,y_1,y_2)  \nonumber\\
&+ {w^*} [E(u_2,u_3,u_1)+E(u_3,u_1,u_2)+\tilde{E}(y_2,y_3,y_1)-\tilde{E}(y_3,y_1,y_2)]\Big\} \,.
\label{PVmrk}
\end{align}
The first factor in this relation takes into account the change from (cosmic) BDS-like normalization back to BDS normalization. Moreover, all transcendental functions should be understood as having been first analytically continued as \eqref{eq:MRK_continuation}, and then evaluated in the limit~\eqref{eq:multi_regge_cross_ratios}. For instance,
\begin{align}
\cE^{(1)} \xrightarrow{\text{MRK}} &-  \frac{1}{2} (\ln^2 u_2+\ln^2 u_3) - 2 \pi i \ln(1-u_1) + 10 \zeta_2 \nonumber \\
&= -\ln^2\tau-\ln^2 |w|-2\pi i\ln\frac{\tau|1+w|^2}{|w|}+10\zeta_2
\end{align}
in this limit, where we also used the shorthand $ww^*\equiv |w|^2$ etc.

The amplitude is invariant with respect to the discrete $y_2\leftrightarrow 1/y_3$ transformation, known as target-projectile symmetry. The latter is a combination of a dihedral flip ($y_2\leftrightarrow y_3$ or $Z_i\to Z_{3-i}$) and a parity ($y_i\to 1/y_i$ or $Z_i\to Z_{i+3}$) transformation. In the multi-Regge limit it becomes equivalent to the inversion $w \to 1/w$, $w^* \to 1/w^*$. Thus if our ansatz for $E,\tilde E$ already respects target-projectile symmetry in general kinematics, we need only compute the first line of \eqn{PVmrk}, and the second line follows from the first by replacing $w\to 1/w$.

The behavior of the amplitude in multi-Regge kinematics may be studied directly within the BFKL approach~\cite{Bartels:2008ce,Bartels:2008sc,Fadin:2011we,Lipatov:2012gk,Dixon:2014iba}, yielding dispersion relation-type integrals for the amplitude. These integrals have the form,
\begin{align}
e^{{\cal R}_6+i\pi\delta}|_{\textrm{MRK}}
 =& \cos\pi\omega_{ab} + i  g^2 \sum_{n=-\infty}^\infty
(-1)^n\left(\frac{w}{\ws}\right)^{\frac{n}{2}}
\cP\int_{-\infty}^{+\infty}
d\nu\frac{\Phi_{\textrm{Reg}}(\nu,n)}{(\nu^2+\frac{n^2}{4})}
\, \nonumber\\
&\qquad\qquad\qquad\qquad\times|w|^{2i\nu}e^{-(\ln \tau+i \pi)\omega(\nu,n)} \, , \nonumber\\
\cP^{(4444)}_{\textrm{NMHV}}\times e^{{\cal R}_6+i\pi\delta}|_{\textrm{MRK}} =& \cos\pi\omega_{ab} +i  g^2 \sum_{n=-\infty}^\infty
(-1)^n\left(\frac{w}{\ws}\right)^{\frac{n}{2}}
\cP\int_{-\infty}^{+\infty}d\nu\frac{\Phi_{\textrm{Reg}}(\nu,n)}{(\nu^2+\frac{n^2}{4})}\bar H(\nu,n)
\, \nonumber\\
&\qquad\qquad\qquad\qquad\times|w|^{2i\nu}e^{-(\ln \tau+i \pi)\omega(\nu,n)}\,,
\label{extMRK}
\end{align}
where
\bea
\omega_{ab} &=& \frac{1}{4}\,\Gamma_{\rm cusp}(g^2)\,\ln|w|^2\,,\\
\delta &=& \frac{1}{4}\,\Gamma_{\rm cusp}(g^2)\,\ln\frac{|w|^2}{|1+w|^4}\,,\\
\tau &=& \sqrt{u_2 u_3}=\frac{(1-u_1)|w|}{|1+w|^2} \,,
\eea
and $\Gamma_{\rm cusp}$ is the cusp anomalous dimension, given in \eqn{Gcusp}. The remaining building blocks of the integrand are known as the 
BFKL eigenvalue $\omega$, the regularized (MHV) impact factor $\Phi_{\textrm{Reg}}$ and the NMHV form factor $\bar H$.\footnote{The NMHV form factor is essentially the ratio between the regularized MHV and NMHV impact factors. In the notations of refs.~\cite{Dixon:2014iba,Dixon:2015iva}, $\Phi^{\textrm{MHV}}_{\textrm{Reg}}=\Phi_{\textrm{Reg}}$ and $\Phi^{\textrm{NMHV}}_{\textrm{Reg}}=\frac{2i\nu+n}{2i\nu-n}\bar H\Phi_{\textrm{Reg}}$.} They were first obtained to the first few orders in perturbation theory by analyzing the effective particles whose exchange dominates in the limit, also by studying the limits of the amplitudes~\cite{Dixon:2014voa}, and more recently to all loops with the help of integrability~\cite{Basso:2014pla}. We refer to the latter reference for their expansion to arbitrary loop order, where in the notation there
\be
\bar H(\nu,n)=\frac{x[u(\nu)+i n/2]}{x[u(\nu)-i n/2]}\,.
\ee
Finally, the $\cP$ on the right-hand side of \eqn{extMRK} denotes the Cauchy principal value of the integral, which has a pole at $n=\nu=0$. Equivalently, we take half the value of the corresponding residue, when evaluating the integral using Cauchy's residue theorem.

Given that all building blocks in the integrand \eqref{extMRK} start at $\cO(1)$, except for $\omega(\nu,n)$ which starts at $\cO(g^2)$, it is easy to see that at $L$ loops ${\cal R}_6^{(L)}$ and $\cP_{\textrm{NMHV}}$ can be written in the limit as a polynomial in $\ln(1-u_1)$, or $\ln\tau$, of degree $L-1$ (plus power-suppressed terms). The coefficient of $\ln^{L-1}\tau$ is referred to as the leading-log (LL) contribution, and similarly the coefficient of $\ln^{L-1-k}\tau$ corresponds to the (next-to)$^k$-leading logarithmic (N$^k$LL) contribution. We provide the result of evaluating the Fourier-Mellin integral in \eqn{extMRK}, separated by logarithmic order, in the ancillary file \texttt{hexMRKL1-7.m}.

Aside from the rational prefactors $1/(1+w^*)$ and $w^*/(1+w^*)$ in the NMHV case, the coefficients of the large logarithms are single-valued harmonic polylogarithms (SVHPLs) in $w,\ws$~\cite{BrownSVHPLs,Dixon:2012yy}.  From \eqns{wdef}{yMRK} it can be seen that these functions have the symbol alphabet $\{w,1+w,w^*,1+w^*\}$, and that the first entry can be only $ww^*$ or $(1+w)(1+w^*)$.  SVHPLs have the important property that they can be uniquely reconstructed from their holomorphic part, defined to be their value at $w^*=0$, also removing any divergent $\ln w^*$.  This observation significantly simplifies the computation of the dispersion integrals~\eqref{extMRK}, since the holomorphic part comes only from the residues satisfying $i\nu= -n/2$~\cite{Drummond:2015jea,Broedel:2015nfp} (see also ref.~\cite{DelDuca:2016lad}).

We can match this holomorphic part directly to the holomorphic part of the multi-Regge limit of our ansatz for the amplitude.  In order to obtain this limit, we iteratively compute the multi-Regge limit of each function in our basis through weight 11.  This can be done using the coproduct entries of these functions, which encode the derivative of each function with respect to $w$. (The multi-Regge limit of all the functions appearing in these coproduct entries is known from the previous iteration in the computation.)  Each function's derivative can thus be matched to a basis of SVHPLs of the appropriate weight.  (There are 3259 SVHPLs at weight 11, including MZVs multiplied by lower weight SVHPLs.) We fix the constants of integration by first computing all functions on the $2\to4$ self-crossing line where $(u,v,w)\to (1-\delta,v,v)$, as described in section 3 of ref.~\cite{Dixon:2016epj} and section~\ref{sec:SelfCrossing} of this paper, and then sending $v\to0$; this limit intersects the $w\to-1$ limit of the multi-Regge limit.  In the coproduct representation of our ansatz, matching to holomorphic data means that we only need to consider the final symbol letters $w$ and $1+w$.  At seven loops, matching at the level of weight 11 functions means that we compare to the $\{11,1,1,1\}$ coproduct component.  This leaves three integration constants unfixed.  However, the constants can be determined as described above, using the self-crossing limit of just the coproduct entries of the amplitude itself, rather than for all weight 12, 13, and 14 functions in $\Hhex$.  (An analogous procedure was used to evaluate the strict collinear and near-collinear limits, where the boundary conditions are also fixed using the same self-crossing line, but this time on the Euclidean sheet where it intersects the collinear limit at its soft endpoint.)

Since $\omega(\nu,n)$ is zero at tree level, any term containing a square or higher power of $\ln\tau$ is determined entirely by lower-loop data~\cite{Caron-Huot:2016owq}. That is, at any given loop order such terms may be predicted using only lower-loop amplitudes and the structure of \eqn{extMRK}, without the need for prior knowledge of the precise form of the integrand building blocks from, say, integrability.  If all parameters in our ansatz can be fixed by such terms, then our bootstrap is in some sense ``pure'', in that it does not rely on external data.

Quite interestingly, while the multi-Regge limits at the $\ln^2\tau$ level fixed all remaining parameters through five loops~\cite{Caron-Huot:2016owq}, they no longer do so at six or seven loops.  The NMHV amplitude can be fixed at six loops using just this data, but in the MHV amplitude a single parameter evades determination in the multi-Regge limit at both six and seven loops, even when the independent predictions of ref.~\cite{Basso:2014pla} are used for the most subleading large logarithms $\ln\tau$ and $\ln^0\tau$.  We will analyze the function accompanying this final six-loop parameter in more detail in subsection \ref{zsec}; however, its appearance is not a problem.  As we now describe, another limit can fix this coefficient. This limit can, in principle, supply an infinite amount of boundary data for the amplitude.

\subsection{Near-collinear operator product expansion}
\label{subsec:ope}

Independent predictions for the behavior of the amplitude in an expansion around the collinear limit may be obtained within the framework of the Wilson loop (or Pentagon) Operator Product Expansion \cite{Alday:2010ku,Basso:2013vsa,Basso:2013aha,Basso:2014koa,Basso:2014jfa,Basso:2014nra,Belitsky:2014sla,Belitsky:2014lta,Basso:2014hfa,Belitsky:2015efa,Basso:2015rta,Basso:2015uxa,Belitsky:2016vyq}. The limit in question is most conveniently described in terms of variables $T=e^{-\tau}$, $S=e^{\sigma}$, and $F=e^{i\phi}$ that appear in the following parametrization of the momentum twistors $Z_i$~\cite{Basso:2013aha},
\be
\left(
\begin{array}{cccccc}
 \frac{S}{\sqrt{F}} & 1 & -1 & 0 & 0 & 0 \\
 0 & 0 & 0 & 1 & 1 & \frac{1}{\sqrt{F} S} \\
 \frac{\sqrt{F}}{T} & 0 & 0 & -1 & 0 & \frac{\sqrt{F}}{T} \\
 \sqrt{F} T & 0 & 1 & 1 & 0 & 0 \\
\end{array}
\right)  \,,
\ee
where the limit corresponds to $\tau\to\infty$ or $T\to0$.
We may express any conformally invariant cross ratio in terms of $T$, $S$, and $F$ with the help of the above parametrization, and the definition of the cross ratio in terms of four-brackets \eqref{eq:fourbrak}. For example, the cross ratios $u$, $v$, and $w$ evaluate to
\be
\begin{aligned}
u\ =&\ \frac{F}{F+F S^2+S T+F^2 S T+F T^2}\,,\\
v\ =&\
\frac{F S^2}{\left(1+T^2\right) \left(F+F S^2+S T+F^2 S T+F T^2\right)}\,,\\
w\ =&\ \frac{T^2}{1+T^2}\, ,
\end{aligned}
\ee
in this parametrization.

As its name suggests, the natural object in the Wilson loop OPE approach is not the amplitude per se, but the closely related, appropriately normalized (framed) Wilson loop. Focusing momentarily on the MHV case, the relation between $\mathcal{E}$ and the corresponding Wilson loop $\mathcal{W}$ is
\be\label{eq:WtoE}
\mathcal{W}=\rho \, \mathcal{E} \, \exp \left({\frac{1}{4}\Gcusp[X-\mathcal{E}^{(1)}]} \right)\,,
\ee
where $\mathcal{E}^{(1)} $ is defined in \eqn{E1}, and $X$ is given by~\cite{Basso:2013aha}
\be
X=-\text{Li}_2(1-u)-\text{Li}_2(1-v)+\text{Li}_2(w)+\ln ^2(1-w)-\ln (1-w) \ln \left(\frac{v}{u}\right)-\ln u \ln v+\frac{\pi^2}{6}\,.\label{eq:XBSV}
\ee
The normalization factor $\rho$ appears in~\eqn{eq:WtoE} to compensate for the fact the Wilson loop is related to the original BDS-like normalized amplitude, not its cosmically normalized cousin. Notice that unlike $\mathcal{E}$, the function $X$ and thus also $\mathcal{W}$ are not dihedrally symmetric. The asymmetry is a consequence of the framing of the hexagonal Wilson loop, which requires dividing it by two pentagonal Wilson loops and multiplying back by a quadrilateral one; the orientation of these auxiliary polygons breaks the dihedral symmetry.

The collinear limit corresponds to $T\to0$, and the Wilson loop OPE is nothing but an expansion of $\mathcal{W}$ in powers $T^{E_i}$ for different exponents $E_i$, which govern how fast the limit is approached. Both the exponent and the coefficient of these powers can be computed exactly in the coupling, as a consequence of the physical interpretation of each term as an excitation of an integrable flux tube, formed by the Wilson lines of the quadrilateral in the aforementioned framing of the Wilson loop.

At loop order $L$ in the weak coupling regime, which is the focus of this paper, the OPE framework predicts that
\be\label{eq:WLOPEexp}
\mathcal{W}^{(L)}=\sum_{m=1}^\infty \sum_{n=-m}^m T^m F^n f^{(L)}_{m,n}(T,S)\,,
\ee
where $f^{(L)}_{m,n}(T,S)$ is a polynomial (of degree $L-1$ for MHV) in $\ln T$, whose coefficients are sums of rational times transcendental functions of $S$. In the above, $m$ and $n$ correspond respectively to the total particle number and helicity of the flux tube excitation, both of which are good quantum numbers.

Here, we will mostly consider the excitations with $m=|n|$, and $n=\pm 1$ or $\pm 2$,
whose contributions to the near-collinear limit have the form,
\be
f_{|n|,n}(T,S)=\sum_{L=1}^\infty g^{2L} f^{(L)}_{|n|,n}(T,S)\,.
\ee
Given that the field content of the theory consists of scalars, fermions and gluons with helicity 0, $\pm 1/2$ and $\pm 1$ respectively, these contributions only come from gluonic excitations of the same helicity. It can be shown that gluons of the same helicity may also form bound states \cite{Basso:2010in,Basso:2014koa}.  The flux tube description of these contributions is
\be\label{eq:gluonOPE}
\begin{aligned}
f_{1,\pm 1}&=\mathcal{W}_{[\pm 1]}\,,\\
f_{2,\pm 2}&=\mathcal{W}_{[\pm 2]}+\mathcal{W}_{[\pm 1,\pm 1]}\,,
\end{aligned}
\ee
where \cite{Basso:2014nra}
\be\label{eq:Wa}
\mathcal{W}_{[a]}=\int \frac{du}{2\pi}\mu_a(u) T^{\gamma_a(u)} S^{ip_a(u)}\,
\ee
denotes the contribution of a single-particle excitation which is a bound state of $|a|$ gluons with helicity $a$ and rapidity $u$, and
\be\label{eq:Wab}
\mathcal{W}_{[a,b]}=\frac{1}{1+
\delta_{ab}}\int \frac{du dv}{(2\pi)^2}\frac{\mu_a(u)\mu_b(v)}{P_{a|b}(u|v)P_{b|a}(v|u)} T^{\gamma_a(u)+\gamma_b(v)} S^{i[p_a(u)+p_b(v)]}\,
\ee
is a superposition of two such excitations, consisting of gluons with helicities $a$ and $b$ such that $ab>0$, and rapidities $u$ and $v$, respectively.

The physical quantities $p_a$ and $\gamma_a=E_a-|a|$ are the momentum and quantum correction to the energy of the flux tube excitation, while $\mu_a$ and $P_{a|b}$ are the measure and pentagon transition, respectively. The finite-coupling expressions for all these quantities have been provided in ref.~\cite{Basso:2014nra}, together with a method for their systematic weak-coupling expansion. In order to avoid clutter, here we illustrate only the leading terms for $ab>0$:
\begin{align}
p_a(u)&=2u+2ig^2 \bigl[\psi(\tfrac{|a|}{2}+iu)-\psi(\tfrac{|a|}{2}-iu)\bigr] + \mathcal{O}(g^4) \,, \\
\gamma_a(u)&=2g^2\bigl[\psi(1+\tfrac{|a|}{2}+iu)+\psi(1+\tfrac{|a|}{2}-iu)-2\psi(1)\bigr] + \mathcal{O}(g^4)\,,\label{weak_E}\\
\mu_a(u)&=(-1)^a g^2\frac{\Gamma(\frac{|a|}{2}+iu)\Gamma(\frac{|a|}{2}-iu)}{\Gamma (|a|)(u+\frac{i|a|}{2})(u-\frac{i|a|}{2})}+\mathcal{O}(g^4) \,,\\
P_{a|b}(u|v)&=\frac{(-1)^b (\tfrac{|a|}{2}-iu) (\tfrac{|b|}{2}+iv) \Gamma(\frac{|a|-|b|}{2}+iu-iv) \Gamma(\frac{|a|+|b|}{2}-iu+iv)}{g^2 \Gamma(\frac{|a|}{2}+iu) \Gamma(\frac{|b|}{2}-iv) \Gamma(1+\frac{|a|-|b|}{2}-iu+iv)} +\mathcal{O}(g^0) \,. \label{weak_P}
\end{align}
We provide expressions for these quantities to eight loops in the ancillary file \texttt{WLOPEblocks.m}.

Because the MHV Wilson loop is a bosonic scalar object, it is invariant under parity, which flips the sign of the helicity. This symmetry implies that
\be
f_{|n|,- n}=f_{|n|,n}\,,\,\text{ or }\,\, {\cW}_{[-a]}={\cW}_{[a]}\,,\,\,{\cW}_{[-a,-b]}={\cW}_{[a,b]}\,,
\ee
as can also be seen explicitly in the leading-order expressions above. From the same expressions, it is evident that $\mathcal{W}_{[a]}$ will start at $\cO(g^2)$.  At $L$ loops, it is a polynomial in $\ln T$ of degree $L-1$, whereas $\mathcal{W}_{[a,b]}$ starts at $\cO(g^8)$, and has degree $L-4$. Remarkably, this implies that the single-gluon state has a distinct signature at order $T^2$, since at $L$ loops all $\ln^k T$ contributions to $f_{2,\pm2}$ with $L-1\le k\le L-3$ come purely from it.

A general algorithm for the weak-coupling evaluation of $\mathcal{W}_{[\pm 1]}$ was developed in ref.~\cite{Papathanasiou:2013uoa}, and later extended to $\mathcal{W}_{[\pm 2]}$~\cite{Papathanasiou:2014yva}. In both cases explicit expressions were obtained up to six loops.\footnote{In particular, the relations between our notations, and those used in the aforementioned papers, are $\mathcal{W}_{[\pm 1]}=I_0=\mathcal{W}_F$, $\mathcal{W}_{[\pm 2]}=\mathcal{W}_{DF}$, ${\cW}_{[\pm 1,\pm 1]}=\mathcal{W}_{FF}=\mathcal{W}_{2[1,1]}/2$.}  In the latter paper, ${\cW}_{[\pm 1,\pm 1]}$ was computed at four loops, by observing that although it is represented as a two-fold integral \eqref{eq:Wab}, it can always be reduced into a sum of products of one-fold integrals, to which the method of ref.~\cite{Papathanasiou:2013uoa} is applicable.  Because this property is independent of loop order, it was also used~\cite{Drummond:2015jea} to determine ${\cW}^{(5)}_{[\pm 1,\pm 1]}$; here we have pushed these three OPE contributions to seven loops. We include them in the files \texttt{WL0-6.m} and \texttt{WL7.m}, through six loops and at seven loops respectively.

Thus we have the complete OPE prediction through seven loops for the $TF^{\pm 1}$ and $T^2 F^{\pm 2}$ terms in the $T\to0$ limit of the MHV amplitude $\cE^{(L)}$, after using the conversion \eqref{eq:WtoE}.  At six loops, we compare this prediction to the limit of our ansatz, with the single parameter that survives the multi-Regge limit. Quite interestingly, at six loops we find that the single such parameter is fixed from the $T^2 F^2 \ln^4T$ contribution, namely from the single-gluon bound state alone.  (Recall that at six loops, the two-gluon superposition only contributes to terms with at most two powers of $\ln T$.) We have checked that our answer for the amplitude also agrees with all the remaining $T^2 F^{\pm2}$ terms, including those sensitive to the two-gluon superposition. Our answer for the limit of the amplitude also contains the $T^2 F^{0}$ terms, which we also include in the ancillary file \texttt{WL0-6.m}. These terms have not been checked against the OPE, although of course we expect them to agree.

At seven loops, we proceed in a slightly different fashion.  The combined constraints, up to and including the multi-Regge limit, again leave one parameter free in the ansatz.  Because the near-collinear limit is computationally demanding at seven loops, we instead examine the origin, where $(u,v,w) \to (0,0,0)$. We observe that through six loops the remainder function has only a very mild singularity, quadratic in logarithms of $u,v,w$. (See section~\ref{OriginSubsection}.) On the other hand, the function multiplying the one free parameter is far more singular at the origin, behaving like $\ln^6 u \ln^4 v \ln^4 w\,+\,$cyclic. Therefore we fix its coefficient by requiring only quadratic behavior in the remainder function at seven loops.  We have verified that this choice for the last parameter matches the OPE predictions in the near-collinear limit for all $T^1 F^{\pm1}$ and $T^2 F^{\pm2}$ terms, i.e.~to the same accuracy checked at six loops.\footnote{The Wilson loop evaluation at seven loops and order $T^2 F^{\pm2}\ln^k T$ for $k < 4$ was done as an expansion for large $S$ through ${\cal O}(1/S^{50})$; for all other values of $k$ the comparison was carried out exactly.} We have also investigated the near-collinear limit of the final seven-loop ambiguity, and find that its limit contains terms of order $T^2 F^{\pm2} \, \ln^6 T$, meaning that its coefficient can also be fixed using just the single-gluon bound state. Again, we extract the $T^2 F^0$ terms in the Wilson loop OPE from the amplitude in this limit, and we include them in the ancillary file \texttt{WL7.m}.

Moving on to the NMHV case, we will focus on taking the near-collinear limit of the $(1111)$ component of the superamplitude, which may be expressed in terms of the $E, \tilde E$ coefficient functions as
\begin{align}
E^{(1111)} &= \frac{1}{2} \Biggl[ 
\frac{S \, [E(u,v,w) - \tilde{E}(u,v,w)]}{S+FT}
+ \frac{E(w,u,v) + \tilde{E}(w,u,v)}{(1+T^2)[1+T(FS+T)]} \nonumber\\
&\hskip0.5cm \null
+ \frac{T(FS+T)^3 \, [E(v,w,u) + \tilde{E}(v,w,u)]}
       {[F(1+T(FS+T))] [F(1+S^2)+T(S(1+F^2)+FT)]} \nonumber\\
&\hskip0.5cm \null
+ \frac{TF^3[E(v,w,u) - \tilde{E}(v,w,u)]} 
       {(S+FT)[F+T(S+FT)] [F(1+S^2)+T(S(1+F^2)+FT)]} \nonumber\\
&\hskip0.5cm \null
+ \frac{T^4 \, [E(w,u,v) - \tilde{E}(w,u,v)]}
       {F(1+T^2) [F+T(S+FT)]}
\Biggr]\,.
\label{E1111def}
\end{align}
This equation follows from the BDS-like normalized analog of \eqn{PVform}, after taking into account that the (1111) component of the $R$-invariants, namely the coefficient of the $\chi_1^4$ term in the definition~\eqref{five_bracket_def},\footnote{Note that ref.~\cite{Basso:2013aha} uses $\eta_i$, rather than $\chi_i$, to refer to dual supercoordinates.} is
\begin{align}
(1)&\to 0\,,\quad(2)\to \frac{F^3 T}{(S+F T) \left(F+S T+F T^2\right) \left(F+F S^2+S T+F^2 S T+F T^2\right)}\,,\nonumber\\
(3)&\to \frac{1}{\left(1+T^2\right) \left(1+F S T+T^2\right)}\,,\quad(4)\to \frac{S}{S+F T}\,,\\
(5)&\to \frac{T (F S+T)^3}{F \left(1+F S T+T^2\right) \left(F+F S^2+S T+F^2 S T+F T^2\right)}\,,\nonumber\\ (6)&\to \frac{T^4}{F \left(1+T^2\right) \left(F+S T+F T^2\right)}\nonumber\,,
\label{RinvFST}
\end{align}
when the cross ratios are parametrized by $F$, $S$ and $T$.

The relation of the (1111) superamplitude component of \eqn{E1111def} to the corresponding component of the NMHV super-Wilson loop, the expansion of the latter in the collinear limit, as well as the predictions for the leading and subleading gluonic OPE contributions to this expansion, then follow straightforwardly from their MHV counterparts, eqs.~\eqref{eq:WtoE}--\eqref{eq:Wab}, upon the simple replacement
\be
\cE\to E^{(1111)}\,,\quad \cW_*\to \cW_*^{(1111)}\,,\quad \mu_a(u)\to \mu_a(u) h_a(u)\,,
\ee
where $*$ can denote the full Wilson loop (component), as in eqs.~\eqref{eq:WtoE}--\eqref{eq:WLOPEexp}, or a particular OPE contribution to it, as in eqs.~\eqref{eq:gluonOPE}--\eqref{eq:Wab}.  (Note that the factor of $\rho$ in \eqn{eq:WtoE} is required in the NMHV case too, given how we cosmically normalize $E$ and $\tilde{E}$ in \eqn{E1111def}.) In addition,
\be\label{eq:NMHVFF}
h_a(u)=\left(\frac{x(u+i a/2)x(u-i a/2)}{g^2}\right)^{\text{sign}(a)}\simeq\left(\frac{u^2+\frac{a^2}{4}}{g^2}+\cO(1)\right)^{\text{sign}(a)}
\ee
are NMHV form factors, responsible for creating OPE excitations that are charged under the $R$-symmetry of the theory. They are expressible in terms of so-called Zhukowski variables,
\be
\quad x(u)=\frac{1}{2}\left(u+\sqrt{u^2-4g^2}\right)\simeq u+\cO(g^2)\,,
\ee
for which we have also indicated the choice of branch at weak coupling.

In contrast to the MHV case, the presence of these form factors implies that OPE contributions with opposite helicity will no longer be equal, e.g.~${\cW}^{(1111)}_{[-a]} \ne {\cW}^{(1111)}_{[a]}$. Nevertheless, we may also evaluate them using the existing summation algorithms mentioned earlier in this section, after taking into account two minor differences for the case $a>0$. First, the presence of inverse powers of the coupling in \eqn{eq:NMHVFF} implies that we have to expand the MHV integrand multiplying the form factor to higher order in $g^2$ than the loop order we wish to compute. And second, particularly for ${\cW}^{(1111)}_{[2]}$, it proves simpler to compute the integral by first defining a reduced integrand $f(u)$ to be the original integrand divided by $(u^2+a^2/4)$, then evaluating the integral of the reduced integrand by residues.  From that integral we obtain the result for the original integrand
by differentiating,\footnote{Note that $f(u)$ is independent of $S$, namely we have taken out of the integral any factors of $\ln S$ coming from the weak-coupling expansion of the integrand, so that the differential operator doesn't act on them.}
\be
\int \frac{du}{2\pi} \left(u^2+\frac{a^2}{4}\right)S^{2i u} f(u)=\left(\frac{1}{(2i)^2}\frac{\partial^2}{\partial (\ln S)^2}+\frac{a^2}{4}\right)\int \frac{du}{2\pi} S^{2i u} f(u)\,.
\ee
In this manner, we have also determined the $TF^{\pm 1}$ and $T^2 F^{\pm 2}$ terms in the collinear OPE expansion of the (1111) NMHV super-Wilson loop component to six loops.  All terms computed agreed with the near-collinear limit of the NMHV amplitude,
which is an independent cross check, since all parameters were fixed by the multi-Regge limit in the NMHV case. We include the near-collinear limit of this component of the super-Wilson-loop in the ancillary file \texttt{W1111L0-6.m}, including $T^2 F^0$ terms which have not yet been checked against the OPE.

Table~\ref{tab:full} summarizes the number of parameters left
after imposing the various constraints we have discussed, through six loops.
The left-hand entry in parentheses is the number for the MHV amplitude,
while the right-hand one is NMHV.  ``MRK'' refers to multi-Regge kinematics,
with the number of N's indicating how many logarithms below the leading
logarithms (LL).
There are a couple of question marks in the table at
six loops, or weight 12, having to do with the number of parity-odd
functions that are not allowed by branch cut and other conditions,
even though their symbol is allowed by symbol-level constraints~\cite{CGG}.
The issue arises because we do not yet have a complete
weight 12 basis at function level, but it is unlikely to affect the final
number by more than one.  There are also some asterisks related to
how the constant $\rho$ is fixed.  If $\rho$ were known ahead
of time at a given loop order, the number of parameters would be exactly
the number shown.  But in practice it is not, and so one should add
the number of asterisks indicated. For example, there are two extra parameters
associated with $\rho$ for the six loop MHV amplitude 
after the strict collinear limits are imposed, in addition to 6 other
parameters that are independent of the value of $\rho$.

The table clearly shows the difference between six loop MHV and all other cases
shown, in that one parameter survives all the way through the multi-Regge
limit and the one flux tube ($T^1$) OPE constraints.  The same is true
at seven loops.  We omit the seven-loop MHV
numbers because we constrained
the result somewhat differently, using symbol-level constraints first
and then reapplying the constraints at function level, so the numbers
are not directly comparable.  However, the number of surviving symbol-level
parameters after the strict collinear limit is 17.  This
number is consistent with the general pattern of the number of parameters
rising by roughly a factor of 3 per loop.

\renewcommand{\arraystretch}{1.25}
\begin{table}[!t]
\centering
\def\sp{@{{\,}}}
\begin{tabular}[t]{l\sp\sp c\sp\sp c\sp\sp c\sp\sp c\sp\sp c\sp\sp c\sp\sp c\sp\sp c}
\hline
Constraint                      & $L=1$\, & $L=2$\, & $L=3$\, & $L=4$\, & $L=5$\, & $L=6$
\\\hline
1. $\Hhex$ & 6 & 27 & 105 & 372 & 1214 & 3692?\\\hline\hline
2. Symmetry       & (2,4) & (7,16) & (22,56) &  (66,190) & (197,602)& (567,1795?)\\\hline
3. Final-entry       & (1,1) & (4,3) & (11,6) &  (30,16) & (85,39) & (236,102)\\\hline
4. Collinear       & (0,0) & (0,0) & $(0^*,0^*)$ &  $(0^*,2^*)$ & $(1^{*3},5^{*3})$ & $(6^{*2},17^{*2})$ \\\hline
5. LL MRK       & (0,0) & (0,0) & (0,0) &  (0,0) & $(0^*,0^*)$ & $(1^{*2}$,$2^{*2})$ \\\hline
6. NLL MRK       & (0,0) & (0,0) & (0,0) &  (0,0) & $(0^*,0^*)$ & $(1^*,0^{*2})$ \\\hline
7. NNLL MRK       & (0,0) & (0,0) & (0,0) &  (0,0) & (0,0) & $(1,0^*)$ \\\hline
8. N$^3$LL MRK       & (0,0) & (0,0) & (0,0) &  (0,0) & (0,0) & (1,0) \\\hline
9. Full MRK       & (0,0) & (0,0) & (0,0) &  (0,0) & (0,0) & (1,0) \\\hline
10. $T^1$ OPE       & (0,0) & (0,0) & (0,0) &  (0,0) & (0,0) & (1,0) \\\hline
11. $T^2$ OPE       & (0,0) & (0,0) & (0,0) &  (0,0) & (0,0) & (0,0) \\\hline
\end{tabular}
\caption{Remaining parameters in the ans\"{a}tze for
the (MHV, NMHV) amplitude after each constraint is applied,
at each loop order. The superscript ``$*$'' (``$*n$'') denotes an additional
ambiguity ($n$ ambiguities) which arises only due to lack of knowledge
of the cosmic normalization constant $\rho$ at the given stage. The ``$?$'' indicates an ambiguity about
the number of weight 12 odd functions that are ``dropouts''; they
are allowed at symbol level but not function level. The seven-loop MHV amplitude was constrained in a somewhat different order. As the parameter counts are not directly comparable it is omitted from the table.}
\label{tab:full}
\end{table}

\subsection{The fate of inter-loop relations}
\label{zsec}

Once we arrive at the final expressions for the six-particle MHV and NMHV amplitudes through seven and six loops respectively, following the steps we described in the previous subsections, we move on to examine their properties. Most of the quantitative and qualitative analysis in various points and lines of the space of kinematics will be done in the next section, but let us conclude this section by noting another new feature that first appears at six loops.

Based on empirical observations up to $L=4$, a relationship between the $L$-loop MHV amplitude and the $(L-1)$-loop NMHV amplitude was conjectured~\cite{Dixon:2014iba}. This relation was then confirmed to hold also at $L=5$~\cite{Caron-Huot:2016owq}. In our notation, which essentially coincides with that of the latter reference, the relation takes the form,
\be\label{eq:MHV-NMHVrel}
 g^2\left(2E {-}  \EE\right)
 = \EE^{y_u,y_u}{+} \EE^{y_w,y_w} {-} 3 \EE^{y_v,y_v}{-} \EE^{v,v} {-} \EE^{1-v,v} 
{+} 2 ( \EE^{y_u,y_v} {+} \EE^{y_w,y_v} ) {-} \EE^{y_u,y_w} {-} \EE^{y_w,y_u},
\ee
where $F^{x,y}$ refers to the corresponding component of the $\Delta_{n-2,1,1}$ ``double'' coproduct of the function $F$, and $E = E(u,v,w)$.

Here, we observe that this relation in fact breaks down at six and seven loops. Very interestingly, this breakdown is closely related to the phenomenon described at the end of section \ref{subsec:initial}, namely the appearance of functions in $\Hhex$ that vanish in the near-collinear limits through ${\cal O}(T)$ and in the multi-Regge limits. These functions are precisely the ones multiplying the parameters of our ansatz that remain free after applying the corresponding constraints.  We saw that there exists one such function at weight 12 that contributes to the MHV amplitude, which we can call $Z(u,v,w)$ for concreteness.  A similar function appears at weight 14, which we call $\tilde{Z}(u,v,w)$.  

Both $Z$ and $\tilde{Z}$ are totally symmetric under dihedral $S_3$ transformations.  Moreover, their parity-even $\{2L-1,1\}$ coproducts vanish identically,
\be
Z^u = Z^v = Z^w = Z^{1-u} = Z^{1-v} = Z^{1-w} = 0,
\label{Zevencopvanish}
\ee
so that they are entirely specified by their $y_u$ coproduct.
The $T$ derivative of any function satisfying \eqn{Zevencopvanish}
can be expressed in terms of the coproducts $Z^{y_u}$, $Z^{y_v}$ and $Z^{y_w}$,
with coefficients that are ${\cal O}(T^0)$ as $T \to 0$.  Because $Z$ is
parity even, the $Z^{y_i}$ are parity-odd, and so they vanish like $T^1$ as
$T\to0$ (times powers of $\ln T$). Thus the $T$ derivative of $Z$
vanishes like $T^1\ln^k T$, and so $Z$ itself must vanish like $T^2 \ln^k T$
(provided that $Z$ is not a constant).
In other words, \eqn{Zevencopvanish} alone is enough to ensure that a
parity even function $Z$ is undetectable in the near-collinear limit
at the level of one flux tube excitation. (It is not as clear to us
why such a function
has to vanish in the multi-Regge limit.)  Why didn't such functions
turn up at lower loop orders?  The answer is that within our function space
there are no parity-even solutions to \eqn{Zevencopvanish} until weight 12!

Returning to the connection between the six loop function $Z$ and
the relation~\eqref{eq:MHV-NMHVrel}, we find that we can satisfy this relation if we shift $\cE^{(6)}\to \cE^{(6)}+\alpha Z(u,v,w)$ for some rational number $\alpha$. In other words, the only effect of the MHV-NMHV relation at six loops is to set the coefficient of $Z(u,v,w)$ in $\cE^{(6)}$ to an incorrect value. This is nontrivial, since \eqn{eq:MHV-NMHVrel} could alternatively yield no solution at all. We suspect that its validity through five loops is linked to the non-existence of analogs of $Z(u,v,w)$ at lower weight.  We should also remark that at seven loops it is not possible to solve \eqn{eq:MHV-NMHVrel} solely by shifting $\cE^{(7)}\to \cE^{(7)} + \beta \tilde{Z}(u,v,w)$ for any $\beta$.

Could there be other relations between MHV and NMHV amplitudes and their double coproducts at one loop higher? By surveying our data up to six loops, we find three such relations:
\be
g^2 E=-2 \left( \EE^{b,m_u} + \EE^{b,m_w}\right)+E^{b,b} + 2\left(E^{b,m_u} + E^{b,m_w}\right) - E^{y_v,y_v}
\ee
and its cyclic images. It remains to be seen whether they continue to hold at higher loops, but paralleling our analysis of the breakdown of \eqn{eq:MHV-NMHVrel}, they may be related to the absence or presence of functions that vanish even faster than $Z$ in the near-collinear limit.


\section{Numerics and number theory in kinematic limits}\label{sec:Numerics}

In the ancillary file \texttt{SixGluonAmpsAndCops.m} accompanying this paper, the six- and seven-loop functions $\cE$, $E$, and $\tilde E$ are expressed in terms of iterated $\Delta_{\bullet,1}$ coproduct entries, supplemented by boundary data at the point (1,1,1). This is formally equivalent to providing these functions in the generalized polylogarithm notation~\eqref{eq:G_func_def} but is far more compact. While at five (and lower) loops, both types of expressions have been published~\cite{Caron-Huot:2016owq,SteinmannWebsite}, the polylogarithmic expressions become too unwieldy by weight 12. In fact, it already proves prohibitively difficult to use these expressions to generate numerics at five loops. However, these functions simplify drastically on a number of codimension two and three subspaces. This allows us to more easily probe the analytic and numerical properties of the amplitudes.

In particular, it was observed by several of the authors~\cite{Dixon:2013eka} that the remainder function appears to behave extremely similarly across loop orders for moderate values of the cross ratios ($0\lesssim u\lesssim 1$). This was interpreted~\cite{Dixon:2014voa, Dixon:2015iva} as a signature of the finite radius of convergence of the perturbative series in planar $\cN=4$ SYM, a property that is expected from the absence of instantons and renormalons. It can be explicitly demonstrated for some quantities that are known at finite coupling, such as the cusp anomalous dimension~\cite{Beisert:2006ez}. The small-coupling expansion of the cusp anomalous dimension has a radius of convergence of $1/16$. Stated differently, at high orders the ratio between the coefficients of successive orders in the coupling approaches $-16.$\footnote{The difference from the ratio of $-8$ quoted in refs.~\cite{Dixon:2014voa,Dixon:2015iva} is due to a different normalization of the coupling there.}

From prior observations at lower loops, we expect that for values of $u,v,w$ of order unity, the ratios of successive loop orders for the functions $E$, $\Et$, and $\EE$ should approach a similar radius of convergence, and should do so quite rapidly in the loop expansion.  Later in this section, we will confirm that this behavior continues to hold at six and seven loops by plotting the ratios of successive loop orders.   First we discuss the functions' behavior at the point $(1,1,1)$ and at the origin.

\subsection{The point \texorpdfstring{$(1,1,1)$}{(1,1,1)}}
\label{SubsectionPoint111}

Because the amplitude is smooth throughout the interior of the Euclidean region, the functions $\cE$ and $E$ have finite values at the point $(u,v,w)=(1,1,1)$. (The function $\tilde E$ vanishes at this point due to parity.) At six and seven loops, these functions become
\bea
\EE^{(6)}(1,1,1) &=& - \frac{2273108143}{6219} \zeta_{12}
\nonumber\\ &&\null\hskip0.0cm
  + \frac{260}{3} \Bigl[ 140 \zeta_5 \zeta_7
     - 56 \zeta_2 \zeta_3 \zeta_7 - 10 \zeta_2 (\zeta_5)^2
     - 60 \zeta_4 \zeta_3 \zeta_5 + 49 \zeta_6 (\zeta_3)^2 \Bigr]
\nonumber\\ &&\null\hskip0.0cm
+ 384 \Bigl[ \zeta_2 \zeta_{7,3} + 14 \zeta_2 \zeta_3 \zeta_7
          + 3 \zeta_2 (\zeta_5)^2 - 7 \zeta_6 (\zeta_3)^2 \Bigr]
\nonumber\\ &&\null\hskip0.0cm
+ 120 \Bigl[ 4 \zeta_4 \zeta_{5,3}
     + 20 \zeta_4 \zeta_3 \zeta_5 - 7 \zeta_6 (\zeta_3)^2 \Bigr]
\nonumber\\ &&\null\hskip0.0cm
+ \frac{5392}{3} \Bigl[ \zeta_{9,3} + 27 \zeta_3 \zeta_9
    + 20 \zeta_5 \zeta_7 - 2 \zeta_2 \zeta_3 \zeta_7
    - \zeta_2 (\zeta_5)^2 - 6 \zeta_4 \zeta_3 \zeta_5
    - 5 \zeta_6 (\zeta_3)^2 \Bigr] , 
\label{EXMHVg6_111}\\ \nonumber \\
\cE^{(7)}(1,1,1) &=& \frac{2519177639}{1260}\zeta_{14} + 
 2496 \Bigl[ 2 \zeta_{5, 3, 3} \zeta_3 - 
    2 \zeta_{5, 3, 3, 3}-\zeta_{5, 3} (\zeta_3)^2 \Bigr]
    \nonumber\\ &&\null\hskip0.0cm
     - 87648 \zeta_{9, 5} + 302160 \zeta_{11, 3} - 
 \frac{61024}{3} \zeta_{9, 3} \zeta_2 - 3264 \zeta_{7, 3} \zeta_4 - 
 7160 \zeta_{5, 3} \zeta_6
\nonumber\\ &&\null\hskip0.0cm
  + \frac{361720}{3} \zeta_8 (\zeta_3)^2 + 
 416 \zeta_2 (\zeta_3)^4 + 206216 \zeta_6 \zeta_3 \zeta_5 - 
 4160 (\zeta_3)^3 \zeta_5 + 95136 \zeta_4 (\zeta_5)^2 
\nonumber\\ &&\null\hskip0.0cm 
 + 
 203136 \zeta_4 \zeta_3 \zeta_7 - 77408 \zeta_2 \zeta_5 \zeta_7 + 
 1208712 (\zeta_7)^2 + 252640 \zeta_2 \zeta_3 \zeta_9 + 
 1082048 \zeta_5 \zeta_9 
\nonumber\\ &&\null\hskip0.0cm  
 - 1241760 \zeta_3 \zeta_{11}
\, ,
\label{EXMHVg7_111}\\ \nonumber \\
E^{(6)}(1,1,1) &=& \frac{5066300219}{6219} \zeta_{12}
\nonumber\\ &&\null\hskip0.0cm
- \frac{344}{3} \Bigl[ 140 \zeta_5 \zeta_7
     - 56 \zeta_2 \zeta_3 \zeta_7 - 10 \zeta_2 (\zeta_5)^2
     - 60 \zeta_4 \zeta_3 \zeta_5 + 49 \zeta_6 (\zeta_3)^2 \Bigr]
\nonumber\\ &&\null\hskip0.0cm
- 528 \Bigl[ \zeta_2 \zeta_{7,3} + 14 \zeta_2 \zeta_3 \zeta_7
     + 3 \zeta_2 (\zeta_5)^2 - 7 \zeta_6 (\zeta_3)^2 \Bigr]
\nonumber\\ &&\null\hskip0.0cm
+ 60 \Bigl[ 4 \zeta_4 \zeta_{5,3}
    + 20 \zeta_4 \zeta_3 \zeta_5 - 7 \zeta_6 (\zeta_3)^2 \Bigr]
\nonumber\\ &&\null\hskip0.0cm
- \frac{9952}{3} \Bigl[ \zeta_{9,3} + 27 \zeta_3 \zeta_9
    + 20 \zeta_5 \zeta_7 - 2 \zeta_2 \zeta_3 \zeta_7
    - \zeta_2 (\zeta_5)^2 - 6 \zeta_4 \zeta_3 \zeta_5
    - 5 \zeta_6 (\zeta_3)^2 \Bigr] .
\label{EXg6_111}
\eea
Notice that similar linear combinations of MZVs appear in $\EE^{(6)}$ and $E^{(6)}$.  This property reflects consistency with a cosmic Galois coaction principle, given the values of lower-loop amplitudes and their coproducts at this
point.\footnote{A convenient way to verify the coaction principle is by writing the MZVs in terms of an $f$ alphabet~\cite{Brown:2011ik} using the {\sc Maple} program {\sc HyperlogProcedures}~\cite{HyperlogProcedures}.}
This point is discussed in depth in our companion paper~\cite{CGG}.

It is interesting to note that the sum $\Sigma^{(L)} = \EE^{(L)}+E^{(L)}$ has somewhat simpler coefficients at this point:
\bea
\Sigma^{(1)}(1,1,1) &=& - 2 \, \zeta_2 \,,
\label{SigmaXg1_111}\\
\Sigma^{(2)}(1,1,1) &=& 16 \, \zeta_4 \,,
\label{SigmaXg2_111}\\
\Sigma^{(3)}(1,1,1) &=& - \frac{527}{3} \, \zeta_6 \,,
\label{SigmaXg3_111}\\
\Sigma^{(4)}(1,1,1) &=&  \frac{19840}{9} \, \zeta_8 \,,
\label{SigmaXg4_111}\\
\Sigma^{(5)}(1,1,1) &=& - \frac{906587}{30} \, \zeta_{10}
\nonumber\\ &&\null\hskip0.0cm
+ 36 \, \Bigl[ 2 \, \zeta_{7,3} + 28 \, \zeta_3 \, \zeta_7
             + 11 \, (\zeta_5)^2 - 4 \, \zeta_2 \, \zeta_3 \, \zeta_5
             - 6 \, \zeta_4 \, (\zeta_3)^2 \Bigr] \,,
\label{Sigmag5_111}\\
\Sigma^{(6)}(1,1,1) &=& \frac{2793192076}{6219} \, \zeta_{12}
\nonumber\\ &&\null\hskip0.0cm
- 28 \Bigl[ 140 \zeta_5 \zeta_7
     - 56 \zeta_2 \zeta_3 \zeta_7 - 10 \zeta_2 (\zeta_5)^2
     - 60 \zeta_4 \zeta_3 \zeta_5 + 49 \zeta_6 (\zeta_3)^2 \Bigr]
\nonumber\\ &&\null\hskip0.0cm
- 144 \Bigl[ \zeta_2 \zeta_{7,3} + 14 \zeta_2 \zeta_3 \zeta_7
     + 3 \zeta_2 (\zeta_5)^2 - 7 \zeta_6 (\zeta_3)^2 \Bigr]
\nonumber\\ &&\null\hskip0.0cm
+ 180 \Bigl[ 4 \zeta_4 \zeta_{5,3}
    + 20 \zeta_4 \zeta_3 \zeta_5 - 7 \zeta_6 (\zeta_3)^2 \Bigr]
\nonumber\\ &&\null\hskip0.0cm
- 1520 \Bigl[ \zeta_{9,3} + 27 \zeta_3 \zeta_9
    + 20 \zeta_5 \zeta_7 - 2 \zeta_2 \zeta_3 \zeta_7
    - \zeta_2 (\zeta_5)^2 - 6 \zeta_4 \zeta_3 \zeta_5
    - 5 \zeta_6 (\zeta_3)^2 \Bigr] . \ \ \ \ \
\label{Sigmag6_111}
\eea
We currently know of no physical reason why the sum should appear
to be simpler.


\subsection{The origin}
\label{OriginSubsection}

In the limit where $u,v,w$ all approach zero, all hexagon functions
become polynomials in the logarithms $\ln u$, $\ln v$, and $\ln w$.  We can determine the (transcendental) coefficients of this polynomial for the functions $\cE$, $E$, and $\tilde E$ by first considering them
at the point $u=1$ and $v,w \to 0$.  This point is in the collinear limit, so
we know the values of these three functions here.  To translate this information to the 
origin, we integrate the amplitude along the $(u,0,0)$ line,
where $v,w \to 0$ but $u$ is generic. Along this line, all hexagon functions become harmonic polylogarithms in $u$ with coefficients that are polynomials
in $\ln v, \ln w$.  

A generic weight-$2L$ hexagon function gives rise to a 
polynomial of total degree $2L$ in
the logarithms.  However, from the leading OPE behavior
on the double-scaling surface $v\to0$ one can show that for the $L$-loop
amplitude, this polynomial should have a degree of at most $L$ in $\ln v$,
and similarly for the degree in $\ln u$ and in $\ln w$ \cite{Gaiotto:2011dt,Dixon:2011pw}.\footnote{The original analyses focused on amplitudes with a one-loop leading OPE contribution. More generally, it can be shown that if the leading OPE contribution an amplitude receives is at $k$ loops, the highest degree of logarithmic divergence is $L-k$ \cite{DelDuca:2018hrv}, with $k=0$ being of course maximal.}

That is indeed the maximum logarithmic behavior that we find for the even
and odd contributions to the ratio function, $V$ and $\tilde{V}$.
Remarkably, for the remainder function through seven loops
we find much less singular behavior, at most quadratic in the total
power of the logarithms.  Also, linear terms in the logarithms do not
appear.  Because ${\cal R}_6(u,v,w)$ is totally symmetric, and there are
only two totally symmetric quadratic polynomials, we find that
\be
 {\cal R}_6^{(L)}(u,v,w)\ \to\  c_1^{(L)} P_1(u,v,w) + c_2^{(L)} P_2(u,v,w)
 + c_0^{(L)} + {\cal O}(u,v,w)\,,
\label{uvwto0}
\ee
as $(u,v,w) \to (0,0,0)$, where
\bea
P_1(u,v,w) &=& \ln^2 u + \ln^2 v + \ln^2 w
      + \ln u \ln v +\ln v \ln w + \ln w \ln u ,
\label{P_1}\\
P_2(u,v,w) &=& \ln u \ln v +\ln v \ln w + \ln w \ln u .
\label{P_2}
\eea
Here $c_1^{(L)}$ and $c_2^{(L)}$ are zeta values of weight $2L-2$,
while $c_0^{(L)}$ has weight $2L$. The values of these constants from two to seven loops are:
\bea
c_1^{(2)} &=& 0 ,  \label{c_1_2}\\
c_1^{(3)} &=& - \frac{5}{2} \, \zeta_4 \,,  \label{c_1_3}\\
c_1^{(4)} &=& \frac{413}{8} \, \zeta_6 - 2 (\zeta_3)^2 \,,  \label{c_1_4}\\
c_1^{(5)} &=& - \frac{6679}{8} \, \zeta_8 + 12 \, \zeta_2 (\zeta_3)^2
               + 40 \, \zeta_3 \zeta_5 \,,  \label{c_1_5}\\
c_1^{(6)} &=& \frac{2033119}{160} \, \zeta_{10} - 159 \, \zeta_4 (\zeta_3)^2
   - 240 \, \zeta_2 \zeta_3 \zeta_5 - 420 \, \zeta_3 \zeta_7
   - 204 \, (\zeta_5)^2 \,,  \label{c_1_6}\\
c_1^{(7)} &=& - \frac{8404209697}{44224} \, \zeta_{12} + 1620 \, \zeta_6 (\zeta_3)^2 + 3252 \, \zeta_4 \zeta_3 \zeta_5 + 2520 \, \zeta_2 \zeta_3 \zeta_7 \nonumber\\
&&\null\hskip0.0cm + 1224 \, \zeta_2 (\zeta_5)^2 + 4704 \, \zeta_3 \zeta_9 + 4368 \, \zeta_5 \zeta_7 + 20 \, (\zeta_3)^4 \,,  \label{c_1_7}
\eea
\bea
c_2^{(2)} &=& \zeta_2 \,,  \label{c_2_2}\\
c_2^{(3)} &=& - 16 \, \zeta_4 \,,  \label{c_2_3}\\
c_2^{(4)} &=& \frac{781}{4} \, \zeta_6 \,,  \label{c_2_4}\\
c_2^{(5)} &=& - \frac{9753}{4} \, \zeta_8 - 8 \, \zeta_2 (\zeta_3)^2 \,,  \label{c_2_5}\\
c_2^{(6)} &=& \frac{2532489}{80} \, \zeta_{10} + 126 \, \zeta_4 (\zeta_3)^2
   + 160 \, \zeta_2 \zeta_3 \zeta_5 \,,  \label{c_2_6}\\
c_2^{(7)} &=& - \frac{9382873343}{22112} \, \zeta_{12} - 1360 \, \zeta_6 (\zeta_3)^2 - 2568 \, \zeta_4 \zeta_3 \zeta_5 - 1680 \, \zeta_2 \zeta_3 \zeta_7 \nonumber\\
&&\null\hskip0.0cm - 816 \, \zeta_2 (\zeta_5)^2 - 8 \, (\zeta_3)^4 \,,  \label{c_2_7}
\eea
and
\bea
c_0^{(2)} &=& \frac{17}{4} \, \zeta_4 \,,  \label{c_0_2}\\
c_0^{(3)} &=& - \frac{1691}{24} \, \zeta_6 + 2 \, (\zeta_3)^2 \,, \label{c_0_3}\\
c_0^{(4)} &=& \frac{32605}{32} \, \zeta_8 - 18 \, \zeta_2 (\zeta_3)^2
    - 40 \, \zeta_3 \zeta_5 \,,  \label{c_0_4}\\
c_0^{(5)} &=& - \frac{2310937}{160} \, \zeta_{10} + 175 \, \zeta_4 (\zeta_3)^2
   + 360 \, \zeta_2 \zeta_3 \zeta_5
   + 420 \, \zeta_3 \zeta_7 + 228 \, (\zeta_5)^2 \,,  \label{c_0_5}\\
c_0^{(6)} &=& \frac{54491355251}{265344} \, \zeta_{12}
    - \frac{5741}{4} \, \zeta_6 (\zeta_3)^2 - 3620 \, \zeta_4 \zeta_3 \zeta_5
    - 3780 \, \zeta_2 \zeta_3 \zeta_7 \nonumber\\
&&\null\hskip0.0cm
    - 1836 \, \zeta_2 (\zeta_5)^2
    - 4704 \, \zeta_3 \zeta_9 - 5208 \, \zeta_5 \zeta_7
    - 14 \, (\zeta_3)^4 \,, \label{c_0_6}\\
c_0^{(7)} &=& - \frac{3768411721}{1280} \, \zeta_{14} + \frac{52815}{4} \, \zeta_8 (\zeta_3)^2 + 31187 \, \zeta_6 \zeta_3 \zeta_5 + 38850 \, \zeta_4 \zeta_3 \zeta_7 \nonumber\\
&&\null\hskip0.0cm + 18750 \, \zeta_4 (\zeta_5)^2 + 42336 \zeta_2 \zeta_3 \zeta_9 + 39312 \, \zeta_2 \zeta_5 \zeta_7 + 156 \, \zeta_2 (\zeta_3)^4\nonumber\\
&&\null\hskip0.0cm
+ 55440 \, \zeta_3 \zeta_{11} + 61824 \, \zeta_5 \zeta_9 + 31560 \, (\zeta_7)^2 + 560 \, (\zeta_3)^3 \zeta_5\,. \label{c_0_7}
\eea
Notice that the potential terms in $c_2^{(L)}$ that are products of just two odd zeta values (for example, $(\zeta_3)^2$ in $c_2^{(4)}$)
all have vanishing coefficients.

Although we do not have a proof that the remainder function is
at most quadratic in $\ln u, \ln v, \ln w$, there is an expectation in the literature that this is indeed the case \cite{Alday:2009dv}. Assuming that it holds, it's worth noting that this constraint can be used to eliminate the MHV ambiguity arising from the function $Z(u,v,w)$, which is not fixed by the multi-Regge limit, nor by the OPE at the level of one flux-tube excitation.  (These two
constraints are apparently closely related by analyticity.)
In the limit $(u,v,w) \to (0,0,0)$, $Z$ is highly singular,
with up to 12 total powers of logarithms (but at most 4 powers
in any individual logarithm):
\be
Z(u,v,w)\ \sim\ \ln^4 u \ln^4 v \ln^4 w\,.
\label{Zuvwto0}
\ee
So in principle we could have used the quadratic logarithmic behavior of
${\cal R}_6^{(6)}(u,v,w)$ in the $(u,v,w)\to(0,0,0)$ limit,
right after the multi-Regge constraint, to remove the final parameter
multiplying $Z$.  In practice, this {\it is} how we removed the parameter
multiplying the analogous function $\tilde{Z}$ at seven loops,
using the fact that
\be
\tilde{Z}(u,v,w)\ \sim\
\ln^4 u \ln^4 v \ln^4 w ( \ln^2 u + \ln^2 v + \ln^2 w )\,. 
\label{tildeZuvwto0}
\ee
After fixing that last parameter, we then verified the OPE limits.


\subsection{The line \texorpdfstring{$(u,u,1)$}{(u,u,1)}}

Now let us consider the line on which $(u,v,w)=(u,u,1)$. Here, the hexagon symbol alphabet ${\cal S}_\text{hex}$ can be written in terms of just two letters, $\{u,1-u\}$. This means that functions on this line can be written in terms of harmonic polylogarithms~\cite{Remiddi:1999ew} in $u$, with weight vector entries drawn from $\{0,1\}$. Parity-odd functions (most importantly, $\tilde E$) vanish on this line, as $\Delta(u,u,1)=0$. From the symmetries of $E$ and $\EE$, we have
\bea
&E(u,u,1)=E(1,u,u) \, ,\\
&\EE(u,u,1)=\EE(u,1,u)=\EE(1,u,u)\,.
\eea
Thus, only $E(u,u,1)$, $E(u,1,u)$, and $\EE(u,u,1)$ represent independent functions.  In the ancillary file \texttt{SixGluonHPLLines.m}, we provide these functions through six loops (seven loops for $\EE(u,u,1)$).

\begin{figure}
\begin{center}
\includegraphics[width=5.5in]{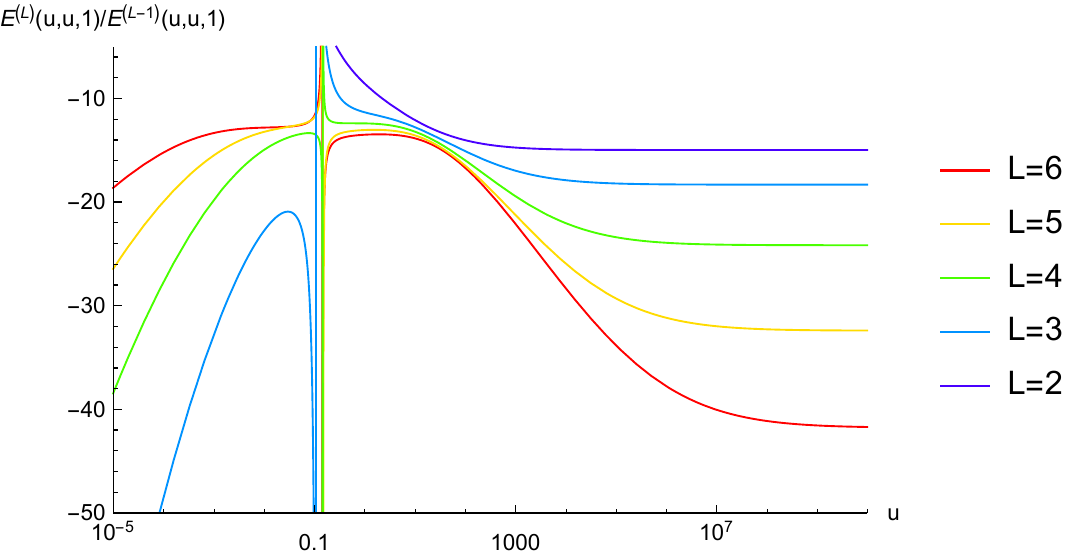}
\end{center}
\caption{$E^{(L)}(u,u,1)/E^{(L-1)}(u,u,1)$ evaluated at successive loop orders. As there are points where $E^{(L-1)}(u,u,1)=0$ in this interval, the plot diverges at those points.}
\label{fig:euu1}
\end{figure}

\begin{figure}
\begin{center}
\includegraphics[width=5.5in]{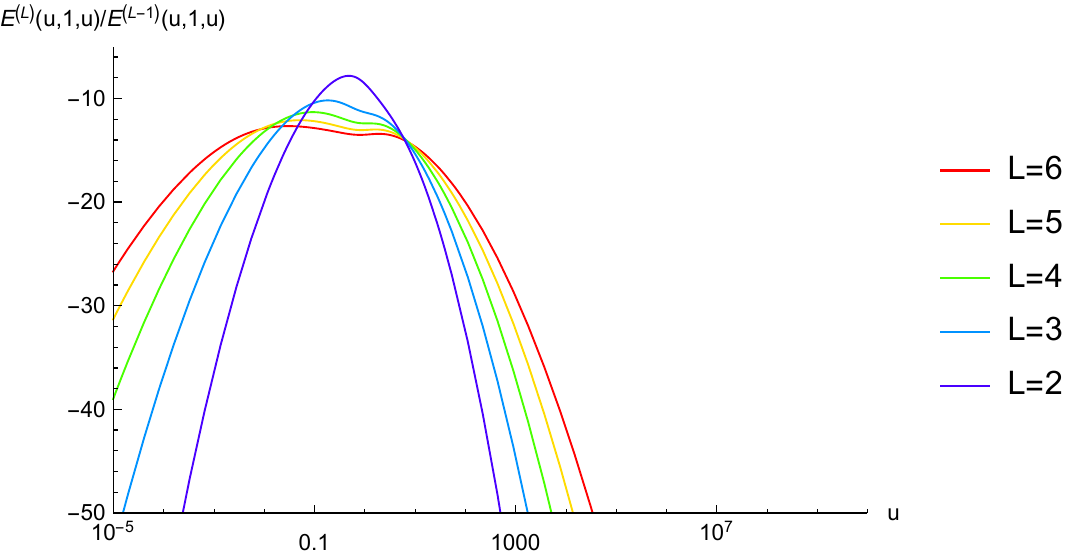}
\end{center}
\caption{$E^{(L)}(u,1,u)/E^{(L-1)}(u,1,u)$ evaluated at successive loop orders.}
\label{fig:eu1u}
\end{figure}

\begin{figure}
\begin{center}
\includegraphics[width=5.5in]{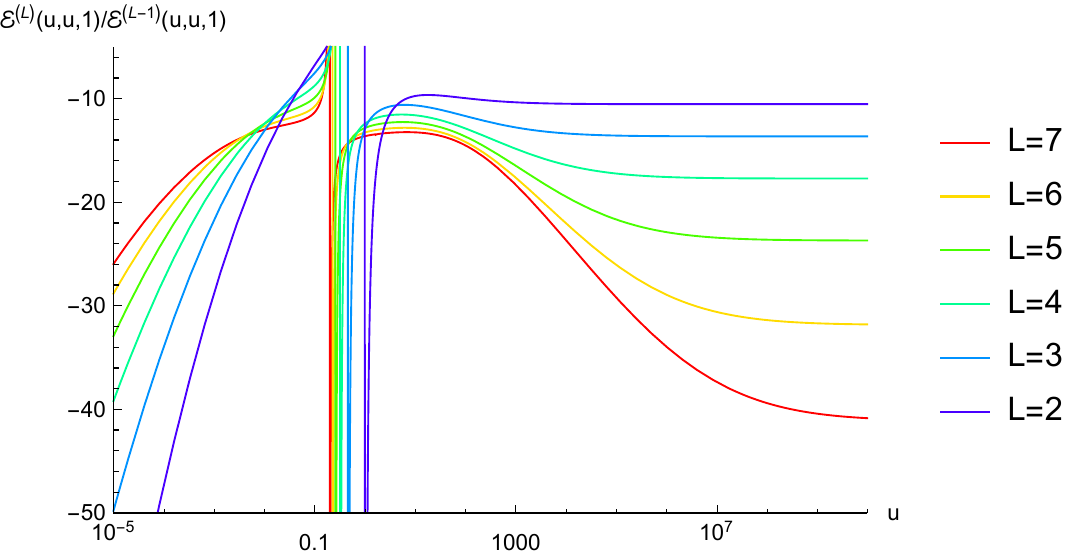}
\end{center}
\caption{$\EE^{(L)}(u,u,1)/\EE^{(L-1)}(u,u,1)$ evaluated at successive loop orders. As there are points where $\EE^{(L-1)}(u,u,1)=0$ in this interval, the plot diverges at those points.}
\label{fig:emuu1}
\end{figure}

We plot $E(u,u,1)$, $E(u,1,u)$, and $\EE(u,u,1)$ in figures \ref{fig:euu1}, \ref{fig:eu1u}, and \ref{fig:emuu1} respectively, through six loops (seven loops for $\EE(u,u,1)$), as ratios of successive loop orders.  For $u$ between $10^{-2}$ and $10^2$, remarkably, the ratios flatten out more and more with each additional loop, and they appear to be steadily approaching the cusp value of $-16$.  Near $u=0.1$ for $E(u,u,1)$, and for $u$ between 0.1 and 1 for $\EE(u,u,1)$ there is a dip/spike feature, which is simply because each function crosses zero at a slightly different value of $u$.  For $u\rightarrow 0$ and $u\rightarrow\infty$, the ratios no longer display the expected radius of convergence, either diverging logarithmically at different rates or, for the examples of $E(u,u,1)$ and $\EE(u,u,1)$ below, approaching constant values that do not have the same ratio of $-16$ between loop orders.

\vfill\eject

\vfill\eject

\vfill\eject

\subsection{The line \texorpdfstring{$(u,1,1)$}{(u,1,1)}}

We can also study the line on which $(u,v,w)=(u,1,1)$. As in the previous subsection, hexagon functions can here be expressed in terms of harmonic polylogarithms. The symmetries of $E$, $\Et$, and $\EE$ give,
\bea
&E(u,1,1)=E(1,1,u), \\
&\Et(u,1,1)=-\Et(1,1,u),\\
&\Et(1,v,1)=0, \\
&\EE(u,1,1)=\EE(1,1,u)=\EE(1,u,1).
\eea
Thus, on this line we can express all functions in terms of $E(u,1,1)$, $E(1,v,1)$, $\Et(u,1,1)$, and $\EE(u,1,1)$. We provide these functions through six loops (seven loops for $\EE(u,1,1)$) in the ancillary file \texttt{SixGluonHPLLines.m}.

\begin{figure}
\begin{center}
\includegraphics[width=5.5in]{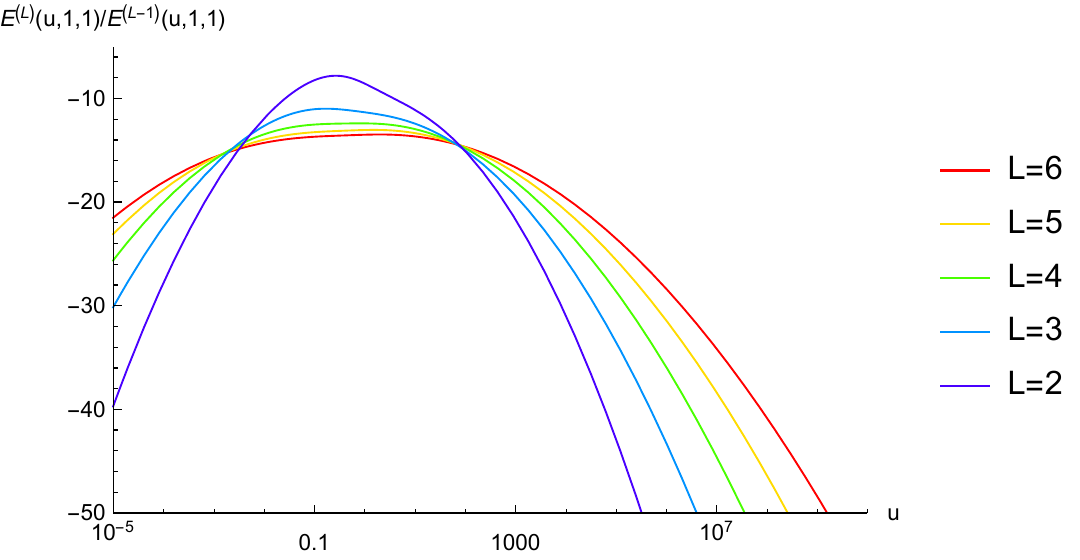}
\end{center}
\caption{$E^{(L)}(u,1,1)/E^{(L-1)}(u,1,1)$ evaluated at successive loop orders.}
\label{fig:eu11}
\end{figure}

\begin{figure}
\begin{center}
\includegraphics[width=5.5in]{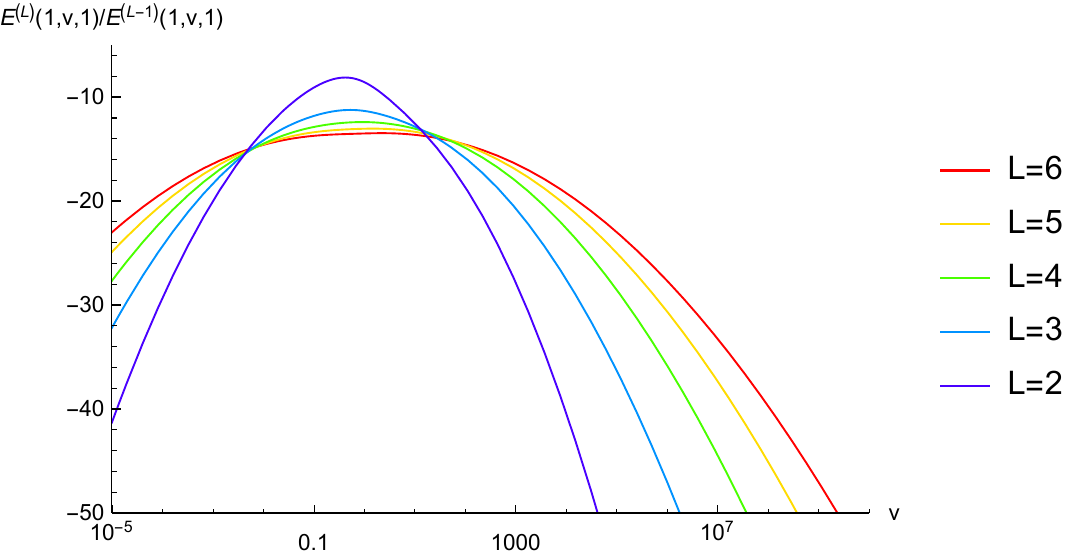}
\end{center}
\caption{$E^{(L)}(1,v,1)/E^{(L-1)}(1,v,1)$, evaluated at successive loop orders.}
\label{fig:e1v1}
\end{figure}

\begin{figure}
\begin{center}
\includegraphics[width=5.5in]{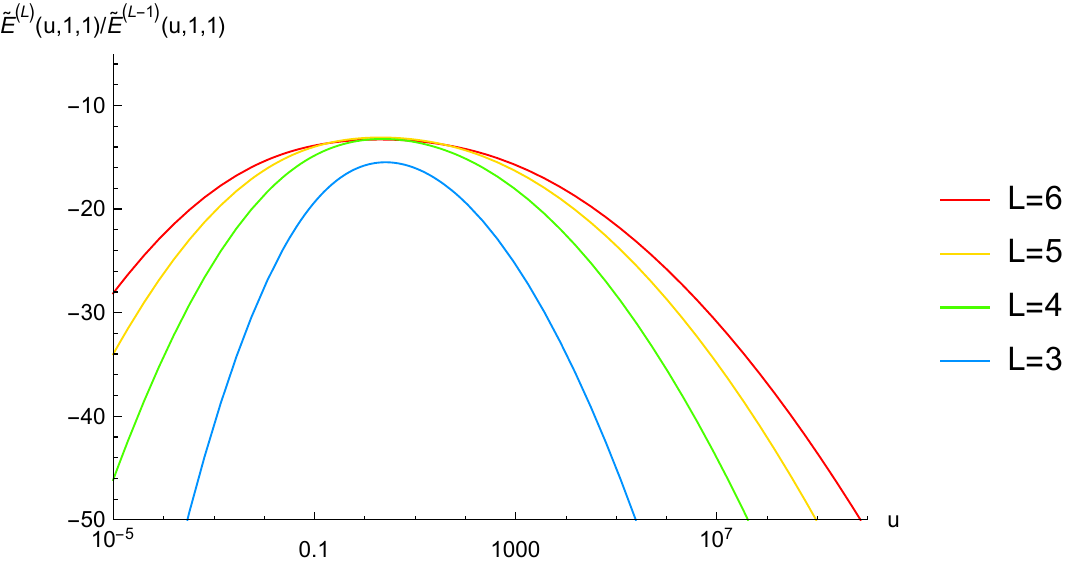}
\end{center}
\caption{$\tilde{E}^{(L)}(u,1,1)/\tilde{E}^{(L-1)}(u,1,1)$ evaluated at successive loop orders.}
\label{fig:etu11}
\end{figure}

\begin{figure}
\begin{center}
\includegraphics[width=5.5in]{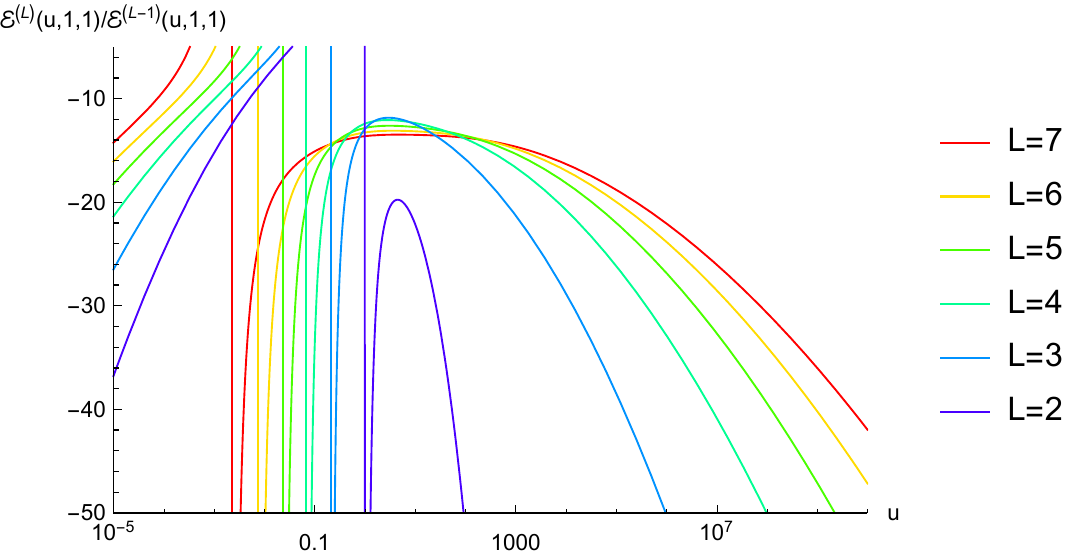}
\end{center}
\caption{$\EE^{(L)}(u,1,1)/\EE^{(L-1)}(u,1,1)$ evaluated at successive loop orders. As there are points where $\EE^{(L-1)}(u,1,1)=0$ in this interval, the plot diverges at those points.}
\label{fig:emu11}
\end{figure}

We plot $E(u,1,1)$, $E(1,v,1)$, $\Et(u,1,1)$, and $\EE(u,1,1)$, in figures \ref{fig:eu11}, \ref{fig:e1v1}, \ref{fig:etu11}, and \ref{fig:emu11} respectively, through six loops (seven loops for $\EE(u,1,1)$), in ratios between successive loop orders. Because $\tilde{E}^{(1)} \equiv 0$, figure \ref{fig:etu11} starts at $L=3$.  Once again the functions have remarkably consistent behavior across loop orders, displaying their rapid approach for $u \sim 1$ to the radius of convergence suggested by the cusp anomalous dimension.

\vfill\eject

\vfill\eject

\vfill\eject

\vfill\eject


\subsection{The line \texorpdfstring{$(u,u,u)$}{(u,u,u)} and strong coupling}

Unlike the previous two lines considered, projecting to the line $u=v=w$ does not reduce our functions to harmonic polylogarithms. Instead, representing them would require cyclotomic polylogarithms~\cite{Ablinger:2011te}. In practice, we evaluate our functions numerically on this line using series expansions around $u=0$, $1$, and $\infty$ with overlapping ranges of convergence.

This line is particularly interesting because the remainder function is known here not only for weak coupling, but for strong coupling as well, due to the work of Alday, Gaiotto, and Maldacena (AGM)~\cite{Alday:2009dv} who represented it in terms of the area of a minimal surface in AdS$_5$.  In fact, the area can be evaluated numerically for generic kinematics, but on this line an analytic strong-coupling formula is available:
\be
R_6^{(\infty)}(u,u,u)\ =\ - \frac{\pi}{6} + \frac{
[ 3\cos^{-1}(1/\sqrt{4u}) ]^2}{3\pi}
- \frac{3}{4} \Li_2\biggl(1-\frac{1}{u}\biggr) - \frac{\pi^2}{12} \,.
\label{R6AGM}
\ee
The last term in \eqn{R6AGM} ensures that $R_6^{(\infty)}(u,v,w)$ vanishes
in the collinear limit.

\begin{figure}
\begin{center}
\includegraphics[width=5.5in]{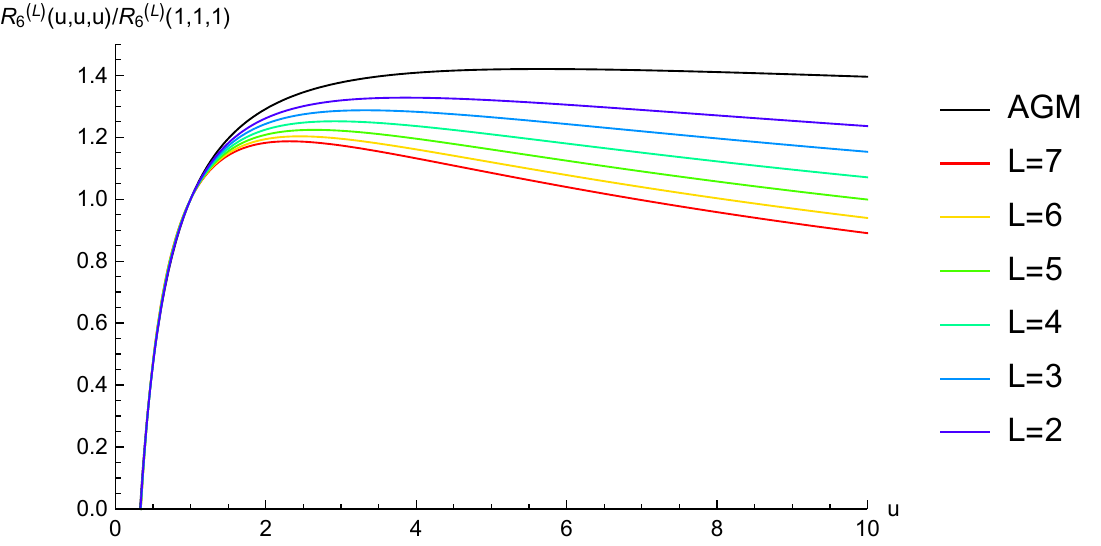}
\end{center}
\caption{Normalized perturbative coefficients of the remainder function, $R_6^{(L)}(u,u,u)/R_6^{(L)}(1,1,1)$, for $L=2$ to 7, plotted along with the strong-coupling result of AGM. The curves all have a remarkably similar shape for $u\lesssim 1$.}
\label{fig:r6uuu}
\end{figure}

In figure \ref{fig:r6uuu}, we plot our results for the remainder function on this line, alongside the strong-coupling AGM result.  We normalize each result by its value at $(1,1,1)$ for ease of comparison.  This normalization forces all curves to go through unity at $u=1$.  Once normalized in this way, the functions are almost indistinguishable for $u<1$, while for $u>1$ their behavior remains rather similar, diverging from each other for large $u$, similarly to their behavior on other lines.

\begin{figure}
\begin{center}
\includegraphics[width=6.0in]{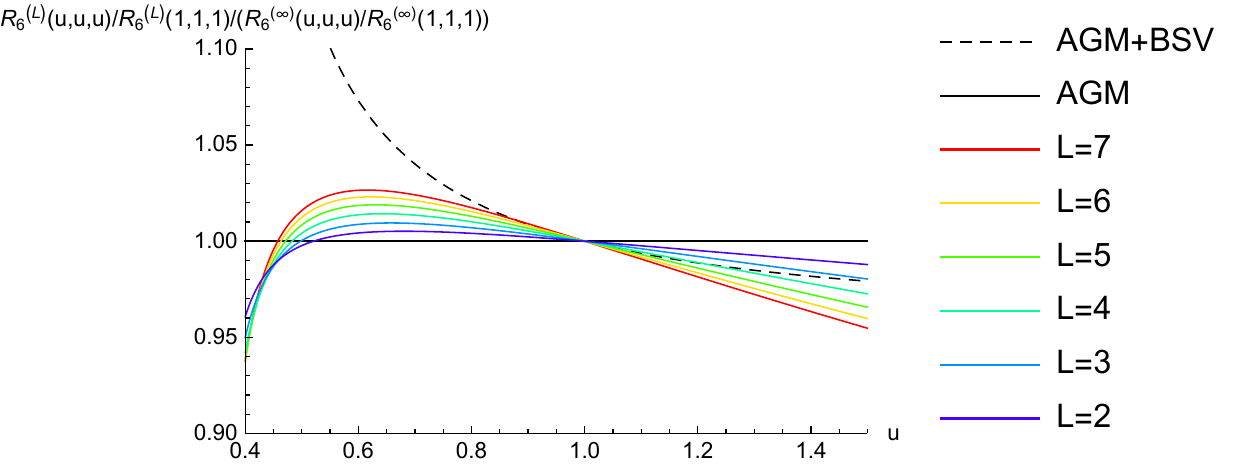}
\end{center}
\caption{For a limited range in $u$, we again plot $R_6^{(L)}(u,u,u)/R_6^{(L)}(1,1,1)$, but now also normalized by the strong-coupling result of AGM.  We also show (dashed line) the strong-coupling result shifted by $-\pi/72$, and normalized by the AGM result.}
\label{fig:r6ratiosuuu}
\end{figure}

In ref.~\cite{Basso:2014jfa} Basso, Sever and Vieira (BSV) pointed out that, because certain scalars become very light at strong coupling, the determinant for fluctuations around the minimal-area surface is parametrically of the same order as the area term; that is, it is also proportional to $g \sim \sqrt{\lambda}$ at strong coupling.  The net result of this computation is to add an additional constant $-\pi/72$ to the AGM result:
\be
R_6^{(\infty)}(u,u,u)\ \to\ R_6^{{\rm (BSV)}}(u,u,u)\
\equiv\ R_6^{(\infty)}(u,u,u) - \frac{\pi}{72} \,.
\label{R6BSV}
\ee
BSV argued that the collinear limit does not commute
with the strong-coupling limit, as a consequence of the constant offset
in \eqn{R6BSV}.  The line $(u,u,u)$ is far from the collinear limit, but
we can still ask, how do large perturbative orders behave,
compared to \eqn{R6AGM} or \eqn{R6BSV}?  To do this, in~\fig{fig:r6ratiosuuu} we re-plot the perturbative results in~\fig{fig:r6uuu}, normalizing everything by $R_6^{(\infty)}(u,u,u)/R_6^{(\infty)}(1,1,1)$ and expanding the scale so we can see differences that were indistinguishable in the previous plot.  The dashed line shows the strong-coupling BSV prediction in the numerator, and the same $R_6^{(\infty)}(u,u,u)/R_6^{(\infty)}(1,1,1)$ in the denominator.  Again all curves must go through unity at $u=1$.  For $0.4 < u < 0.1$, all the perturbative results have the same shape as the strong-coupling result $R_6^{(\infty)}(u,u,u)$ to within a few percent.  However, a naive extrapolation to yet higher loop orders suggests that the shape for $0.5 < u < 0.1$ is starting to resemble the AGM+BSV prediction more closely than the AGM prediction.

\vfill\eject


\section{The self-crossing limit}\label{sec:SelfCrossing}

A hexagonal Wilson loop can be deformed until two of its lines, on opposite sides of the hexagon, almost cross each other.  This limit was first studied for Wilson loops in planar ${\cal N}=4$ super-Yang-Mills theory in
refs.~\cite{Georgiou:2009mp,Dorn:2011gf,Dorn:2011ec}.  In ref.~\cite{Dixon:2016epj}, it was pointed out that the limit for $2\to4$ kinematics mimics double parton scattering in hadronic collisions, and this limit and an analogous limit for $3\to3$ kinematics were thoroughly explored.  Making use of the anomalous dimension matrix for crossing Wilson lines~\cite{Korchemskaya:1994qp}, it was argued that the singular parts of $\EE$ should not depend on the residual kinematic variable characterizing the self-crossing kinematics.  By matching to the multi-Regge limit discussed in section~\ref{subsec:multi-regge}, it was possible to predict the singular terms to high loop orders.  In this section, we will test these predictions and provide results for related but nonsingular terms in the self-crossing limit.

In self-crossing kinematics, the cross ratios $(u,v,w)$ approach $(1-\delta,v,v)$ with $\delta \to 0$, after performing an analytic continuation from the Euclidean region.  In the $2\to4$ case, the analytic continuation to be performed is $u\to e^{-2\pi i}u$, as in \eqn{eq:MRK_continuation}; the
surviving cross ratio $v$ is restricted to $0 < v < 1$, and the small parameter $\delta$ is positive.  In the $3\to3$ case, $v$ is restricted to either $v<0$ or $v>1$, and $\delta$ is negative, so we will express the results in terms of $|\delta|=-\delta$.  For $v<0$, the analytic continuation is $u\to e^{+2\pi i}u$, $v\to e^{+\pi i}v$, $w\to e^{+\pi i}w$.  We can go from $v<0$ to $v>1$ by analytically continuing $\ln(1-v) \to \ln(v-1) - i\pi$.  In fact,
$\EE_{3\to3}(1+|\delta|,v,v)$ is smooth around $v=\pm\infty$, and so no analytic continuation is necessary after writing it in the right representation in terms of harmonic polylogarithms.

The amplitude is simpler in $3\to3$ self-crossing kinematics than in $2\to4$ kinematics.  The reason is that the hexagonal Wilson loop has an alternating structure, incoming-outgoing-incoming-$\cdots$, which makes it quasi-Euclidean.  As a result, all the singularities as $|\delta| \to 0$ are in the imaginary part of the amplitude.  In ref.~\cite{Dixon:2016epj}, a formula was provided for a nonsingularly-framed Wilson loop,
\be
\frac{1}{2\pi i} \frac{d {\cal W}^{\rm ns}_{3\to3}}{d\lnden}\ =\
\exp\Bigl[ - \frac{\Gcusp}{4} ( \lndene{2} - L^2 ) \Bigr]
\, g(\lnden,\Gcusp) \,,
\label{defineg}
\ee
where $L=\ln(1-1/v)$ and the function $g(\lnden,\Gcusp)$
(not to be confused with the coupling constant $g$)
was given through 7 loops.
The relation between ${\cal W}^{\rm ns}$ and $\EE$ was also determined,
except that the $\EE$ in ref.~\cite{Dixon:2016epj} differs from the one here
by the constant factor $\rho$ given in~\eqn{rho}.
Correcting for that difference, we have
\be
\EE_{3\to3}\ =\ \frac{ {\cal W}^{\rm ns}_{3\to3} }{\rho} \times
 \exp\Bigl[ - \frac{\Gcusp}{4} \Bigl( L^2 - 4 \, \zeta_2 \Bigr) \Bigr]
\,. \label{EXfromW}
\ee
Since $\rho$ is a nonsingular constant, the derivative with respect
to $\lnden$ passes right through it, and we obtain,
\be
\frac{1}{2\pi i} \frac{d \EE_{3\to3}}{d\lnden}\ =\
\exp\Bigl[ - \frac{\Gcusp}{4} ( \lndene{2} - 4 \, \zeta_2 ) \Bigr]
\, \frac{g(\lnden,\Gcusp)}{\rho} \,.
\label{EXsing}
\ee
This formula controls all of the singular terms in $\EE_{3\to3}$ as $\delta\to0$.
Note that these terms are totally independent of $v$.

In ref.~\cite{Dixon:2016epj}, the function $g(\lnden,\Gcusp)$
was evaluated completely through 7 loops, and partially at 8 and 9 loops, using
the connection to the multi-Regge limit.  However,
motivated by the fact that the self-crossing limit is a virtual
Sudakov region, where virtual gluons are confined to narrow jets
as $|\delta| \to 0$, it is possible to give another representation
for the singular terms in $\EE_{3\to3}$ in the self-crossing limit:
\bea
\frac{1}{2\pi i} \frac{d \EE_{3\to3}}{d\lnden} &=&
\frac{g^2}{\rho} \exp\Bigl[ \tfrac{1}{2} \zeta_2 \Gcusp + 2 \Gamma_3 \Bigr]
\nn\\
&&\hskip0.0cm\null
\times 2 \int_0^\infty d\nu J_1(2\nu) \exp\Bigl[
 - \frac{1}{4} \Gcusp [\lambda(\nu)]^2 - \Gvirt \lambda(\nu) \Bigr]\,,
\label{gSimon}
\eea
where $J_1$ is the first Bessel function and
\be
\lambda(\nu) = 2 ( \ln\nu + \gamma_E ) - \lnden,
\label{lambdadef}
\ee
with $\gamma_E$ the Euler-Mascheroni constant.
We have also introduced two additional anomalous dimensions,
$\Gvirt$ and $\Gamma_3$, which arise in
the large-rapidity limit of the flux-tube
spectrum~\cite{Beisert:2006ez,Basso:2013vsa,Basso:2013aha,SeverAmps15}.

Define the semi-infinite
matrix~$K$~\cite{Beisert:2006ez} and vector $\kappa^{\rm eff}$
by
\be
K_{ij} = 2 j (-1)^{j(i+1)} 
\int_0^\infty \frac{dt}{t} \frac{J_i(2gt) J_j(2gt)}{e^t-1} \,, \qquad
\kappa_j^{\rm eff} =
\int_0^\infty \frac{dt}{t} \frac{J_j(2gt) J_0(2gt) - gt \delta_{j,1}}{e^t-1} \,,
\label{Kkappa}
\ee
and $M = (1+K)^{-1}$, $Q_{ij} = j (-1)^{j+1} \delta_{ij}$.  Then the three
anomalous dimensions are
\bea
\Gcusp &=& 4 g^2 M_{11} \,,  \label{GcuspfromM}\\
\Gvirt &=& 4 g \sum_{j=1}^\infty M_{1j} \kappa^{\rm eff}_j \,,  \label{GvirtfromM}\\
\Gamma_3 &=& - 2 \sum_{i,j,k=1}^\infty \kappa^{\rm eff}_i Q_{ij} M_{jk} \kappa^{\rm eff}_k 
  + \int_0^\infty \frac{dt}{t} \frac{1 - [J_0(2gt)]^2}{e^t-1}
  - \frac{1}{2} \zeta_2 \Gcusp \,. \label{G3fromM}
\eea
We have
\bea
\Gvirt &=& - 12 \zeta_3 g^4 + (80\zeta_5+16\zeta_2\zeta_3) g^6
-(700\zeta_7+80\zeta_2\zeta_5+168\zeta_3\zeta_4) g^8 + \ldots,
\label{Gvexpand}\\
\Gamma_3 &=& \frac{7}{2}\zeta_4 g^4
-\Bigl(\frac{115}{3}\zeta_6+\frac{56}{3}(\zeta_3)^2 \Bigr) g^6
+ \Bigl(\frac{1701}{4}\zeta_8 + 260\zeta_3\zeta_5
+ 28\zeta_2(\zeta_3)^2 \Bigr) g^8 + \ldots,~~~~
\label{G3expand}
\eea
while the first few terms of $\Gcusp$ are given in \eqn{Gcusp}.

Given these anomalous dimensions, it is straightforward to
evaluate \eqn{gSimon} perturbatively to very high orders
(or indeed, nonperturbatively).  After expanding the integrand
in $g^2$, one encounters integrals of the Bessel function $J_1(2\nu)$
multiplied by powers of $\ln\nu + \gamma_E$.  The integrals can
be performed by Taylor expanding the identity~\cite{BabusciDattoli}
\be
2 \int_0^\infty d\nu J_1(2\nu) e^{2(\ln\nu+\gamma_E)\xi}
\ =\ e^{2\gamma_E\xi} \frac{\Gamma(1+\xi)}{\Gamma(1-\xi)}
\label{BesselIntegral}
\ee
around $\xi=0$.  In an ancillary file, {\tt SelfCrossSingular.m},
we evaluate \eqn{gSimon} through 20 loops. (We have multiplied it back
by $\rho$, because we only know $\rho$ to 7 loops.)

In the rest of this section, we provide some information about
the terms in the MHV\footnote{The NMHV amplitude is trickier
to evaluate in the self-crossing limit because some of the $R$-invariants
diverge there, necessitating a higher-order expansion of the transcendental
functions in $\delta$, which we won't perform here.}
amplitude $\EE$ that are nonsingular as $\delta\to0$.
The limit $v \to \pm\infty$ is smooth, i.e.~there are no $\ln v$ singularities
for $\EE$ in this limit.  We will give the singular
terms, as well as some nonsingular constants, in this limit.  The following
expressions match eqs.~(3.28)--(3.33) of ref.~\cite{Dixon:2016epj}
through five loops, after accounting for
a factor of two difference in the loop expansion parameter, and
(starting from three loops) from the factor of $\rho$ used to
define $\EE$ here:
\bea
{\cal E}_{3\to3}^{(0)}(v=\infty) &=& 1 \,, \label{Evinfty_0}\\
{\cal E}_{3\to3}^{(1)}(v=\infty) &=& 2\pi i \, \lnden \, + \, 2 \zeta_2 \,,
\label{Evinfty_1}\\
{\cal E}_{3\to3}^{(2)}(v=\infty) &=& 2 \pi i \, \biggl[
  - \frac{1}{3} \lndene{3} \, + 2 \zeta_2 \lnden
 \, + \, 8 \zeta_3 \biggr]
  + 28 \zeta_4 \,, \label{Evinfty_2}\\
{\cal E}_{3\to3}^{(3)}(v=\infty) &=& 2 \pi i \, \biggl[ \frac{1}{10} \lndene{5}
 \, - \, 2 \zeta_3 \lndene{2}
 \, + \, 2 \zeta_4 \lnden 
 \, - \, 84 \zeta_5
 \, + \, 36 \zeta_2 \zeta_3 \biggr] \nn\\
&&\hskip0cm\null
   - \frac{3787}{12} \zeta_6
\, + \, 2 (\zeta_3)^2 \,, \label{Evinfty_3}\\
{\cal E}_{3\to3}^{(4)}(v=\infty) &=& 2 \pi i \, \biggl[
  - \frac{1}{42} \lndene{7}
  \, - \, \frac{1}{5} \zeta_2 \lndene{5}
  \, - \, \frac{1}{3} \zeta_3 \lndene{4}
  \, - \, \frac{14}{3} \zeta_4 \lndene{3} \nn\\
&&\hskip0.7cm\null
  \, + \, 4 ( 4 \zeta_5 - 3 \zeta_2 \zeta_3 ) \lndene{2} 
  \, - \, \frac{1}{3} \Bigl( 13 \zeta_6 + 72 (\zeta_3)^2 \Bigr) \lnden \nn\\
&&\hskip0.7cm\null
  + 1141 \zeta_7 - 476 \zeta_2 \zeta_5
  - 68 \zeta_3 \zeta_4  \biggr] \nn\\
&&\hskip0.0cm\null
  + \frac{56911}{9} \zeta_8 + 20 \zeta_{5,3} 
  - 92 \zeta_3 \zeta_5 + 272 \zeta_2 (\zeta_3)^2 
\,, \label{Evinfty_4}
\eea
\bea
{\cal E}_{3\to3}^{(5)}(v=\infty) &=& 2 \pi i \, \biggl[ 
  \frac{1}{216} \lndene{9}
  \, + \, \frac{2}{21} \zeta_2 \lndene{7}
  \, + \, \frac{5}{9} \zeta_3 \lndene{6}
  \, + \, \frac{18}{5} \zeta_4 \lndene{5} \nn\\
&&\hskip0.7cm\null
  \, + \, \frac{4}{3} ( 6 \zeta_5 + 7 \zeta_2 \zeta_3 ) \lndene{4}
  \, + \, \frac{4}{9} \Bigl( 115 \zeta_6 + 54 (\zeta_3)^2 \Bigr) \lndene{3}
\nn\\ &&\hskip0.7cm\null
  \, + \, 2 \Bigl( - 55 \zeta_7 + 68 \zeta_2 \zeta_5
                        + 44 \zeta_3 \zeta_4 \Bigr) \lndene{2} \nn\\
&&\hskip0.7cm\null
  + \frac{4}{9} \Bigl( 257 \zeta_8 + 1170 \zeta_3 \zeta_5
                                   - 18 \zeta_2 (\zeta_3)^2 \Bigr) \lnden
\nn\\
&&\hskip0.7cm\null
- \frac{40369}{2} \zeta_9 + 7645 \zeta_2 \zeta_7
+ \frac{3119}{2} \zeta_3 \zeta_6
+ 2295 \zeta_4 \zeta_5 - 184 (\zeta_3)^3 \biggr] \nn\\
&&\hskip0cm\null
- \frac{2668732849}{16800} \zeta_{10}
- \frac{1467}{7} \zeta_{7,3}
+ \frac{4868}{5} \zeta_2 \zeta_{5,3}
- 2851 \zeta_4 (\zeta_3)^2
\nn\\&&\hskip0cm\null
- 12044 \zeta_2 \zeta_3 \zeta_5
+ \frac{5819}{2} \zeta_3 \zeta_7 + \frac{14169}{7} (\zeta_5)^2 \,,
\label{Evinfty_5}
\eea
\bea
{\cal E}_{3\to3}^{(6)}(v=\infty) &=& 2 \pi i \, \biggl[ 
  - \frac{1}{1320} \lndene{11}
  \, - \, \frac{1}{36} \zeta_2 \lndene{9}
  \, - \, \frac{1}{4} \zeta_3 \lndene{8}
  \, - \, \frac{34}{21} \zeta_4 \lndene{7}
\nl && \hskip0.7cm\null
- \frac{1}{15} \Bigl( 152 \zeta_5 + 90 \zeta_2 \zeta_3 \Bigr)
                  \lndene{6}
\, - \, \frac{1}{15} \Bigl( 632 \zeta_6 + 308 (\zeta_3)^2 \Bigr)
                  \lndene{5}
\nl && \hskip0.7cm\null
- \frac{1}{3} \Bigl( 555 \zeta_7 + 444 \zeta_2 \zeta_5
                              + 428 \zeta_3 \zeta_4 \Bigr) \lndene{4}
\nl && \hskip0.7cm\null
- \frac{1}{9} \Bigl( 5446 \zeta_8 + 6648 \zeta_3 \zeta_5
                     + 1272 \zeta_2 (\zeta_3)^2 \Bigr) \lndene{3}
\nl && \hskip0.7cm\null
- \Bigl( -168 \zeta_9+1860 \zeta_2 \zeta_7+1336 \zeta_4 \zeta_5
     +\frac{3184}{3} \zeta_6 \zeta_3+128 \zeta_3^3 \Bigr) \lndene{2}
\nl && \hskip0.7cm\null
- \Bigl( \frac{26441}{15} \zeta_{10} + 496 \zeta_2 \zeta_3 \zeta_5
   + 744 \zeta_4 (\zeta_3)^2 + 6464 \zeta_3 \zeta_7
   + 3312 (\zeta_5)^2 \Bigr) \lnden 
\nl && \hskip0.7cm\null
+\frac{98955281}{160} \zeta_{11} - \frac{8892}{5} \zeta_{5,3,3}
+ \frac{72}{5} \zeta_3 \zeta_{5,3}
- \frac{4301861}{18} \zeta_2 \zeta_9 - \frac{1647589}{20} \zeta_4 \zeta_7
\nl && \hskip0.7cm\null
- \frac{160435}{3} \zeta_6 \zeta_5 - \frac{984359}{30} \zeta_8 \zeta_3
- 1400 \zeta_2 (\zeta_3)^3 + 11520 (\zeta_3)^2 \zeta_5 \biggr]
\nl && \hskip0.0cm\null
+ \frac{1993553577827}{398016} \zeta_{12}
- 176 \zeta_{6,4,1,1} + 1818 \zeta_{9,3}
- \frac{157183}{14} \zeta_2 \zeta_{7,3} - 10272 \zeta_4 \zeta_{5,3}
\nl && \hskip0.0cm\null
+ 335729 \zeta_2 \zeta_3 \zeta_7 + \frac{3000535}{14} \zeta_2 (\zeta_5)^2
+ 180706 \zeta_4 \zeta_3 \zeta_5 + 85268 \zeta_6 (\zeta_3)^2
\nl && \hskip0.0cm\null
- \frac{981071}{9} \zeta_3 \zeta_9 - 151629 \zeta_5 \zeta_7 - 452 (\zeta_3)^4
\,,
\label{Evinfty_6}
\eea
and
\bea
{\cal E}_{3\to3}^{(7)}(v=\infty) &=& 2 \pi i \, \biggl[
\frac{1}{9360} \lndene{13} + \frac{1}{165} \zeta_2 \lndene{11}
+ \frac{13}{180} \zeta_3 \lndene{10}
+ \frac{55}{108} \zeta_4 \lndene{9}
\nl && \hskip0.7cm\null
+ \frac{1}{15} \Bigl( 74 \zeta_5 + 40 \zeta_2 \zeta_3 \Bigr) \lndene{8}
+ \frac{1}{63}  \Bigl( 1265 \zeta_6 + 764 (\zeta_3)^2 \Bigr) \lndene{7}
\nl && \hskip0.7cm\null
+ \Bigl( \frac{617}{3} \zeta_7 + \frac{572}{5} \zeta_2 \zeta_5
       + \frac{1030}{9} \zeta_3 \zeta_4 \Bigr) \lndene{6}
\nl && \hskip0.7cm\null
+ \Bigl( \frac{4546}{9} \zeta_8 + \frac{11884}{15} \zeta_3 \zeta_5
       + \frac{1036}{5} \zeta_2 (\zeta_3)^2 \Bigr) \lndene{5}
\nl && \hskip0.7cm\null
+ \Bigl( 4396 \zeta_9 + 2740 \zeta_2 \zeta_7 + 2592 \zeta_4 \zeta_5
  + \frac{16390}{9} \zeta_6 \zeta_3 + 384 (\zeta_3)^3 \Bigr) \lndene{4}
\nl && \hskip0.7cm\null
+ \Bigl( \frac{66892}{9} \zeta_{10} + \frac{17792}{3} \zeta_2 \zeta_3 \zeta_5
  + \frac{9712}{3} \zeta_4 (\zeta_3)^2 + \frac{39904}{3} \zeta_3 \zeta_7
\nl && \hskip1.4cm\null
  + \frac{20000}{3} (\zeta_5)^2 \Bigr)
  \lndene{3}
\nl && \hskip0.7cm\null
+ \Bigl( 21672 \zeta_{11} + 31584 \zeta_2 \zeta_9 + 24188 \zeta_4 \zeta_7
  + \frac{50828}{3} \zeta_6 \zeta_5 + \frac{37160}{3} \zeta_8 \zeta_3
\nl && \hskip1.4cm\null
  + 2192 \zeta_2 (\zeta_3)^3 + 9232 (\zeta_3)^2 \zeta_5 \Bigr) \lndene{2}
\nl && \hskip0.7cm\null
+ \Bigl( \frac{157796480}{6219} \zeta_{12} + 19072 \zeta_2 \zeta_3 \zeta_7
  + 9120 \zeta_2 (\zeta_5)^2 + 28720 \zeta_4 \zeta_3 \zeta_5
\nl && \hskip1.4cm\null
  + 9628 \zeta_6 (\zeta_3)^2 + 97440 \zeta_3 \zeta_9 + 96480 \zeta_5 \zeta_7
  + 1376 (\zeta_3)^4 \Bigr) \lnden
\nl && \hskip0.7cm\null
- \frac{323782470527}{16800} \zeta_{13} + \frac{153703}{7} \zeta_{7,3,3}
- \frac{3165856}{175} \zeta_{5,5,3}
- \frac{98576}{5} \zeta_2 \zeta_{5,3,3}
\nl && \hskip0.7cm\null
+ \frac{10966}{7} \zeta_3 \zeta_{7,3}
+ \frac{235982}{5} \zeta_5 \zeta_{5,3}
+ \frac{5336}{5} \zeta_2 \zeta_3 \zeta_{5,3}
\nl && \hskip0.7cm\null
+ \frac{1912832287}{240} \zeta_2 \zeta_{11}
+ \frac{48297353}{56} \zeta_4 \zeta_9
+ \frac{1116988937}{560} \zeta_6 \zeta_7
\nl && \hskip0.7cm\null
+ \frac{20103611}{12} \zeta_8 \zeta_5
+ \frac{8268675281}{8400} \zeta_{10} \zeta_3
+ 115968 \zeta_2 (\zeta_3)^2 \zeta_5
\nl && \hskip0.7cm\null
+ \frac{78950}{3} \zeta_4 (\zeta_3)^3
- 332586 (\zeta_3)^2 \zeta_7
- \frac{2649110}{7} \zeta_3 (\zeta_5)^2 \biggr]
\nl  && \hskip0.0cm\null
- \frac{26743565967068063}{119750400} \zeta_{14}
+ \frac{407994591437}{142560} \zeta_{11,3}
- \frac{15377712919}{19440} \zeta_{9,5}
\nl && \hskip0.0cm\null
- \frac{231544}{5} \zeta_{5,3,3,3}
+ \frac{3473431}{6} \zeta_2 \zeta_{9,3}
+ \frac{397712}{3} \zeta_2 \zeta_{6,4,1,1}
+ 2384 \zeta_3 \zeta_{5,3,3}
\nl && \hskip0.0cm\null
- 835791 \zeta_4 \zeta_{7,3}
- \frac{358802}{15} \zeta_6 \zeta_{5,3}
+ \frac{9804}{5} (\zeta_3)^2 \zeta_{5,3}
- \frac{277080440}{27} \zeta_2 \zeta_3 \zeta_9
\nl && \hskip0.0cm\null
- \frac{52133752}{3} \zeta_2 \zeta_5 \zeta_7
- \frac{486350}{9} \zeta_2 (\zeta_3)^4
- \frac{118926569}{12} \zeta_4 \zeta_3 \zeta_7
- \frac{63595231}{12} \zeta_4 (\zeta_5)^2
\nl && \hskip0.0cm\null
- 6455307 \zeta_6 \zeta_3 \zeta_5
- \frac{472744763}{180} \zeta_8 (\zeta_3)^2
+ \frac{68075555}{16} \zeta_3 \zeta_{11}
\nl && \hskip0.7cm\null
+ \frac{555617147837}{23760} \zeta_5 \zeta_9
+ \frac{5061150659977}{285120} (\zeta_7)^2
+ \frac{187276}{3} (\zeta_3)^3 \zeta_5
\,.
\label{Evinfty_7}
\eea
The $\lnden$ terms in the results are all consistent with
the predictions of \eqns{EXsing}{gSimon}.

As the point $v=1$ is approached from above, i.e.~from the $3\to3$ side,
the results are also relatively simple.  Because the $\lnden$-dependent
terms are identical to those presented above for $v=\infty$,
here we give only the finite terms:
\bea
{\cal E}_{3\to3}^{(0),{\rm fin}}(v\to1^+) &=& 1 \,, \label{Eveq1p_0}\\
{\cal E}_{3\to3}^{(1),{\rm fin}}(v\to1^+) &=& 0 \,, \label{Eveq1p_1}\\
{\cal E}_{3\to3}^{(2),{\rm fin}}(v\to1^+) &=&
2 \pi i \times 4 \zeta_3 - 10 \zeta_4 \,, \label{Eveq1p_2}\\
{\cal E}_{3\to3}^{(3),{\rm fin}}(v\to1^+) &=&
2 \pi i \Bigl[ - 32 \zeta_5 + 16 \zeta_2 \zeta_3 \Bigr]
+ \frac{35}{3} \zeta_6 \,, \label{Eveq1p_3}\\
{\cal E}_{3\to3}^{(4),{\rm fin}}(v\to1^+) &=&
2 \pi i \biggl[ 312 \zeta_7
  - 152 \zeta_2 \zeta_5 + 20 \zeta_3 \zeta_4 \biggr]
\nn\\  &&\hskip0.0cm\null
- \frac{77}{3} \zeta_8 + 24 \zeta_{5,3} 
+ 120 \zeta_3 \zeta_5 + 120 \zeta_2 (\zeta_3)^2 
\,,~~ \label{Eveq1p_4}\\
{\cal E}_{3\to3}^{(5),{\rm fin}}(v\to1^+) &=&
2 \pi i \biggl[
  - 3428 \zeta_9 + 1640 \zeta_2 \zeta_7 - 360 \zeta_4 \zeta_5
  - \frac{598}{3} \zeta_6 \zeta_3 - 48 (\zeta_3)^3 \biggr]
\nn\\  &&\hskip0.0cm\null
   - \frac{3961}{3} \zeta_{10} - 192 \zeta_{7,3} + 240 \zeta_2 \zeta_{5,3}
   - 1920 \zeta_2 \zeta_3 \zeta_5 + 576 \zeta_4 (\zeta_3)^2
\nn\\  &&\hskip0.0cm\null
   - 2688 \zeta_3 \zeta_7 - 1356 (\zeta_5)^2
\,, \label{Eveq1p_5}\\
{\cal E}_{3\to3}^{(6),{\rm fin}}(v\to1^+ ) &=&
2 \pi i \biggl[
37972 \zeta_{11} + 32 \zeta_{5,3,3} - 32 \zeta_3 \zeta_{5,3}
 - \frac{53464}{3} \zeta_2 \zeta_9 + 5580 \zeta_4 \zeta_7
\nn\\  &&\hskip0.7cm\null
 + \frac{13508}{3} \zeta_6 \zeta_5
 + \frac{53338}{9} \zeta_8 \zeta_3
 - \frac{896}{3} \zeta_2 (\zeta_3)^3 + 1040 (\zeta_3)^2 \zeta_5 \biggr]
\nn\\  &&\hskip0.0cm\null
 + \frac{167184257}{6219} \zeta_{12}
 + \frac{5392}{3} \zeta_{9,3} - 1920 \zeta_2 \zeta_{7,3} + 480 \zeta_4 \zeta_{5,3}
\nn\\  &&\hskip0.0cm\null
 + 15360 \zeta_2 \zeta_3 \zeta_7
 + 7560 \zeta_2 (\zeta_5)^2
 - 13584 \zeta_4 \zeta_3 \zeta_5 - 4740 \zeta_6 (\zeta_3)^2
\nn\\  &&\hskip0.0cm\null
 + 48528 \zeta_3 \zeta_9 + 48080 \zeta_5 \zeta_7
\,, \label{Eveq1p_6}\\
{\cal E}_{3\to3}^{(7),{\rm fin}}(v\to1^+) &=&
2 \pi i \biggl[ - 426820 \zeta_{13}
 - 576 \zeta_{7,3,3} + 576 \zeta_{5,5,3} + 480 \zeta_3 \zeta_{7,3}
 + 64 \zeta_5 \zeta_{5,3}
\nn\\  &&\hskip0.7cm\null
 - 64 \zeta_2 \zeta_3 \zeta_{5,3}
 + 203640 \zeta_2 \zeta_{11} - \frac{261296}{3} \zeta_4 \zeta_9
 - 83366 \zeta_6 \zeta_7
\nn\\  &&\hskip0.7cm\null
 - \frac{1061720}{9} \zeta_8 \zeta_5
 - \frac{2072438}{15} \zeta_{10} \zeta_3
 + 13280 \zeta_2 (\zeta_3)^2 \zeta_5 + \frac{6560}{3} \zeta_4 (\zeta_3)^3
\nn\\  &&\hskip0.7cm\null
 - 2656 (\zeta_3)^2 \zeta_7 - 2464 \zeta_3 (\zeta_5)^2
\biggr]
\nn\\  &&\hskip0.0cm\null
 - \frac{4788480727}{1260} \zeta_{14}
 - 4992 \zeta_{5,3,3,3} + 302160 \zeta_{11,3} - 87648 \zeta_{9,5}
 + \frac{5600}{3} \zeta_2 \zeta_{9,3}
\nn\\  &&\hskip0.0cm\null
 + 4992 \zeta_3 \zeta_{5,3,3} - 3264 \zeta_4 \zeta_{7,3}
 + 5944 \zeta_6 \zeta_{5,3} - 2496 (\zeta_3)^2 \zeta_{5,3}
 + 84640 \zeta_2 \zeta_3 \zeta_9
\nn\\  &&\hskip0.0cm\null
 - 212000 \zeta_2 \zeta_5 \zeta_7
 + 203136 \zeta_4 \zeta_3 \zeta_7 + 95136 \zeta_4 (\zeta_5)^2
 + 158840 \zeta_6 \zeta_3 \zeta_5
\nn\\  &&\hskip0.0cm\null
 + \frac{258520}{3} \zeta_8 (\zeta_3)^2
 - 2080 \zeta_2 (\zeta_3)^4 - 1241760 \zeta_3 \zeta_{11}
 + 1082048 \zeta_5 \zeta_9
\nn\\  &&\hskip0.0cm\null
 + 1208712 (\zeta_7)^2 - 4160 (\zeta_3)^3 \zeta_5 
\,. \label{Eveq1p_7}
\eea
Notice that the expressions in the limit $v\to1^+$ are quite a bit
simpler than those in the $v\to\infty$ limit.

There do not seem to be many restrictions on the MZVs
that can appear at $v = \pm\infty$, or in the (singular) imaginary
part as $v\to 1^+$.  However, the real part as $v\to 1^+$, which is finite
as $\de\to0$, does have MZV restrictions, and a coaction
principle operates at this point as well, even though it is not on
the Euclidean sheet.  In fact, the dimensionality of the space of
allowed MZVs here is exactly the same as for the Euclidean
point $(1,1,1)$ through at least weight 11,
although the actual values are different~\cite{CGG}.

We have also evaluated the MHV amplitude as a function of $v$
for the $3\to3$ configurations with $v<0$ and $v>1$, and the $2\to4$
configuration with $0<v<1$, through seven loops, in terms
of harmonic polylogarithms.
Because these expressions are rather lengthy at six and seven loops,
where they go beyond ref.~\cite{Dixon:2016epj}, we provide them
in an ancillary file, {\tt SelfCross.m}.

\vfill\eject


\section{Conclusions}\label{sec:Conclusions}

In this work, we have bootstrapped the six- and seven-loop six-particle MHV amplitudes, as well as the six-loop NMHV amplitude, in planar ${\cal N}=4$ super-Yang-Mills theory. The space of functions $\Hhex$ that provides the initial ansatz for this procedure is presented in a companion paper~\cite{CGG}. This space of functions takes advantage of constraints following from cosmic Galois theory and the extended Steinmann relations, and appears to be the minimal space required to express the six-point amplitudes in this theory.

In order to fit the amplitude into this space, we have absorbed several constants into a function of the coupling $\rho(g^2)$. This function is universal for MHV and NMHV, and it is natural to wonder whether it has physical meaning. Its number-theoretic content appears to be related to the cusp anomalous dimension, but further work will be necessary to establish whether the relation between these quantities can be made precise. Such work might hint at a physical meaning and answer the question:  why is it possible to normalize these amplitudes in such a way that they respect a cosmic Galois coaction principle between loop orders?

Collinear and final-entry constraints both powerfully constrain the ansatz formed out of this space of functions. Together with information from the multi-Regge limit, they {\it almost} suffice to determine the amplitude completely. A novel feature we encountered, however, was the appearance of one potential contribution to the MHV amplitude at six loops, and again one at seven loops, that vanished in each of these limits. We successfully detected these contributions by their nonvanishing terms in the near-collinear expansion to next-to-leading order, which is governed by the Pentagon Operator Product Expansion, and in particular its first gluon bound state. It is interesting to note that while the multi-Regge limit can be reached from the OPE by an analytic continuation that also includes this state \cite{Basso:2014pla}, it is nevertheless weaker. In other words, the analytic continuation causes some information to be lost.

Since the Pentagon OPE offers an infinite amount of boundary information, the presence of functions that cannot be detected by their multi-Regge and strict collinear limits is in principle not a problem at higher loops. In practice, accessing increasingly subleading terms in the near-collinear expansion also becomes increasingly difficult. It corresponds to acting with more and more derivatives on the transcendental functions, which in turn requires tabulating many complicated expressions.  An alternative would be to partially resum a relatively simple subset of OPE excitations, that spans a more general limit where the amplitude does not vanish when evaluated there in a strict sense. The double scaling limit~\cite{Basso:2014nra} is an excellent example.  It corresponds to sending one cross ratio to zero while holding the other two cross ratios fixed. The resummation of the infinite set of purely gluonic excitations that dominate this limit was initiated in ref.~\cite{Drummond:2015jea}. Indeed, the distinct signature of OPE excitations that were resummed in the latter paper, if pursued to six and seven loops, would be able to fix the final free parameters there.

Better yet, it would be very exciting if we could harness the power of the OPE in order to access the origin in cross ratio space, where we saw that functions which can only be detected from the OPE have unusually divergent behavior, in contrast to the remainder function, which is only quadratic in the logarithms of the cross ratios through seven loops. While the origin is outside of the radius of convergence of the OPE, it certainly lies within the subspace spanned by the aforementioned double scaling limit. We are optimistic that it can be reached by means of an analytic continuation, as was the case with the multi-Regge limit.

We also studied the self-crossing limit in this paper.  The singular terms as this limit is approached were computed to seven loops, and found to match previous predictions~\cite{Dixon:2016epj}.  A compact, Sudakov-based formula for the singular terms was presented that can be evaluated to very high loop order.  In addition we studied the nonsingular terms, which at one point are governed by the coaction
principle in a nontrivial way.

We continue to observe that the amplitudes have very numerically consistent ratios between successive loop orders.  These ratios approach the radius of convergence previously observed for the cusp anomalous dimension. It is interesting that the ratios of loop orders are already beginning to converge at this order, which is quite atypical for other examples of quantum field theories with finite radii of convergence.

Finally, we are optimistic that these techniques can be applied to other theories without such a high degree of symmetry.

\vskip0.5cm
\noindent {\large\bf Acknowledgments}
\vskip0.3cm

\noindent We are grateful to Francis Brown, Erik Panzer, Oliver Schnetz
and Mark Spradlin for many illuminating discussions.
This research was supported in part by the National Science Foundation under
Grant No.\ NSF PHY17-48958, by the US Department of Energy under contract
DE--AC02--76SF00515, the Munich Institute for Astro- and Particle Physics (MIAPP) of the DFG cluster of excellence ``Origin and Structure of the Universe'', the Perimeter Institute for Theoretical Physics, the Danish National Research Foundation (DNRF91), a grant from the Villum Fonden, a Starting Grant \mbox{(No.\ 757978)} from the European Research Council, a grant from the Simons Foundation (341344, LA), the European Union's Horizon 2020 research and innovation program under grant agreement \mbox{No.\ 793151}, and a Carlsberg Postdoctoral Fellowship (CF18-0641). SCH's research is supported by the National Science and Engineering Council of Canada.
LD thanks Perimeter Institute, the Simons Foundation, the Hausdorff Institute for Mathematics, and Humboldt University Berlin for hospitality during this project.  AM is grateful to the Higgs Centre at U.~Edinburgh for hospitality, and LD, MvH, AM, and GP thank the Kavli Institute for Theoretical Physics and the Galileo Galilei Institute for hospitality.  Research at Perimeter Institute is supported by the Government of Canada through Industry Canada and by the Province of Ontario through the Ministry of Economic Development and Innovation.

\noindent 



\providecommand{\href}[2]{#2}\begingroup\raggedright\endgroup

\end{fmffile}
\end{document}